\documentclass[preprint,showpacs,preprintnumbers,amsmath,amssymb]{revtex4-1}
\usepackage{graphicx}
\usepackage{bm}
\usepackage{color}
\usepackage{amssymb}
\usepackage{epsfig}
\DeclareGraphicsExtensions{.png,.pdf}

\begin{document}

\title{Entanglement dynamics of multi-parametric random states: a single parametric formulation}
\author{Devanshu Shekhar and Pragya Shukla}
\affiliation{ Department of Physics, Indian Institute of Technology, Kharagpur-721302, West Bengal, India }
\date{\today}

\widetext

\begin{abstract}
	
A non-ergodic quantum state of a many body system is in general random as well as multi-parametric, former due to a lack of exact information due to complexity and latter reflecting its varied behavior in different parts of the Hilbert space. An appropriate representation for the reduced density matrix of such a state is a  generalized, multi-parametric Wishart ensemble with unit trace. Our theoretical analysis  of these ensembles not only  resolves the controversy about the growth rates of the  average information entropies
of the generic states but also leads to new insights in their entanglement dynamics.  While the state itself is multi-parametric, we find that the growth of  the average measures can be described in terms of  an information-theoretic function, referred  as the complexity parameter. The latter in turn leads to a common mathematical formulation of the measures for a wide range of states; it could also act as a possible tool for hierarchical arrangement of the entangled states of different systems.

\end{abstract}

\maketitle

\section{Introduction}

Notwithstanding extensive previous studies  of the  bipartite entanglement  of
many body quantum states  \cite{zhh, nahum1, vhbr, nh, bgl, ha,  zl, jhn, rpv, efg, imptlrg, rpk27, rpk28, rpk15, rpk19, rpk24, rpk25, rpk26, nhkw, rpk14, rpk17, rpk20, rpk21, rpk23, puz,  hp, ss, arul,   rpk12, rpk13,  rpk16, rpk22}, many basic issues remain still unsolved or elude unanimity. One such issue is the entanglement dynamics of a typical, non-ergodic quantum state of a many body system, although undeniably relevant for realistic applications but lacking deeper insights so far. 
The standard route to determine the entanglement measures of  bipartite quantum states is through the reduced density matrix of the subsystems; the latter can be well modeled by stationary Wishart random matrix ensembles if the system is in an  ergodic state \cite{zhh, apsk}. But, contrary to ergodicity, non-ergodicity of a state is system specific and can depend on many system parameters. Indeed a typical many body state need not be ergodic and can exhibit different degree of localization e.g. localized, partially localized or extended as well based on system parameters.
A density matrix representation of such a state, in any physically motivated basis, is further complicated by a lack of exact information due to underlying complexity that leaves it beyond the scope of a stationary Wishart random matrix ensemble. A  theoretical formulation of the entanglement measures of such  states  requires, therefore, considerations of many different multi-parametric Wishart ensembles where system dependence manifests through statistical diversity of the matrix elements. But with mathematical analysis often based on system specific approximations dictated by existing parametric ranges, the results available for a typical state of one system need not be applicable to another one (of same or different system); this further adds to its technical complexity. Nonetheless the omnipresence of such states in realistic applications renders their analysis unavoidable. One way to address the issue is to seek a common mathematical formulation, if possible,  for the entanglement measures of the multi-parameteric Wishart ensembles representing  a wide range of many body states and is primary focus of the present study.           

A maximally entangled state of a  many body quantum system is expected, by definition, to be ergodic; the latter corresponds to an equal a-priori probability in all basis states such that the information the knowledge of the state can provide about its subunits is minimum. This aspect renders the state a  huge potential applicability in quantum information processes and  has led to intense research seeking  its existence/engineering during past two decades. A typical many body state is however expected to evolve in time and need not retain its initial entanglement characteristics. As the  number of ways for a generic state to be entangled are more than unentangled, the entanglement measures in general tend to grow in time with a growth rate that can reveal many system specific characteristics e.g. many body localization, the existence of a local conserved charge etc. Indeed the dynamics of entanglement with time is of fundamental interest too: it is believed to be the primary mechanism for the thermalization in quantum systems evolved by unitary evolution. Besides, a  scaling of entanglement entropy (EE) with the system size also contains information about quantum phase transitions and can act as a criteria to distinguish a thermal phase from a many-body localized one \cite{nh}. The studies on black hole entropy and thermalization in many body systems  have also indicated bounds on ''scrambling'' of quantum information i.e the time beyond which quantum information can not be retrieved completely due to our  inability to perform measurements \cite{hp, ss, nhkw,bgl}. The scrambling time ($t_s$), when the information is distributed among all the sub-units, is lower bounded by the logarithm of degrees of freedom in the system, $t_s \geq \log N$.

While the richness of applications, fundamental as well as industrial,  has motivated in past many studies on  the time-dynamics of entanglement \cite{zhh, efg, zl, imptlrg, bgl, arul, puz}, many issues still remain unresolved. For example, it is not yet understood  how the saturation (fast scrambling) is achieved for systems with sparse connectivity, e.g., black-holes \cite{bgl}. Besides the system-conditions can also change over time, thus rendering it  relevant to study the dynamics with changing system parameters directly. This in turn requires information about the eigenvalue dynamics of a  multi-parametric Wishart ensemble with ensemble parameters mimicking  the change of system conditions; the latter formulation was derived recently by one of us \cite{ptche, pche}. In the present study, we apply the formulation  to analyze the behavior of entanglement measures as system conditions  vary.

An analysis of the entanglement growth, over time, of a many body state also requires a comparative study of a specific measure at various stages of system conditions (e.g. those changing with time). Previous research studies have introduced many measures  to characterize the entanglement in a multipartite state.  Unfortunately the criteria/hierarchical arrangement of states with increasing entanglement predicted by different measures lacks unanimity i.e the relative degree of entanglement between two state indicated by one measure differs from other \cite{virmani2000ordering}. Previously it was believed that the issue does not arise in bipartite systems and the degree of entanglement can be consistently described by any of the standard measure e.g. von Neumann entropy, R\'enyi entropy etc. Indeed, based on systems without conservation laws or disorder, previous studies predicted a linear growth of both the entropies \cite{jhn, rpk12, rpk13, rpk14, rpk15, rpk16, rpk17, rpk19, rpk20, rpk21, rpk22, rpk24, rpk25, rpk26, rpk27, rpk28}.
The recent studies however indicate different growth rates of the two entropies with time if the quantum system is subjected  to global physical constraints e.g. symmetry, conservation laws \cite{zl, imptlrg} or local ones e.g. disorder. For example, in systems exhibiting diffusive transport of some conserved quantity e.g spin, charge, energy, the  R\'enyi entropies grow much more slowly in time than the  von Neumann entropy. This leads to a natural query regarding the role of physical constraints  on the entanglement growth. The present work  analyzes the issue in context of one global constraint i.e the states with and without time-reversal symmetry.

The paper is organized as follows. We begin by a brief review of the relevant definitions in section II.  The reduced density matrix of a bipartite quantum state can in general be written as a Wishart matrix. The latter belongs to a random matrix ensemble if the system is complex, e.g, one with complicated interactions. As the non-ergodic state in general depends on many system parameters, this manifests in the ensemble through distribution parameters.  As discussed in section II. A, the  reduced density matrix of interest then belongs to a general class of multiparametric Wishart random matrix ensembles. This in turn affects the distribution of the Schmidt eigenvalues and thereby the average behavior of the entanglement measures. While changing system conditions can lead to a multi-parametric variation of the ensemble, the latter's evolution in the matrix space is governed by the a single functional, referred as the complexity parameter $Y$; this is discussed in section II.B. This is followed by a derivation of $Y$ dependent growth of the ensemble averaged entanglement entropy $R_1$ in section III.A. Contrary to $R_1$, the second-order R\'enyi entropy $R_2$ is an experimentally more accessible measure of entanglement, e.g., by measuring a local operator \cite{ha}. This motivates us to derive $Y$ dependent growth of ensemble averaged $R_2$ in section III.B along with an interpretation of $Y$ in terms of time $t$ in section III.C. The details of a numerical verification of our theoretical predictions is presented in section IV. 
As the complexity parameter is an important backbone of our analysis, it is necessary to analyze its implications in context of entanglement dynamics; 
this is discussed in section V  followed by a review of our main results and open questions in section VI. To avoid defocussing, the steps related to some of the derivations are moved from main text to the appendices.

\section{Component matrix ensemble for a pure state}

The density operator $\rho$ for  a pure state $|\Psi\rangle$ is expressed as $\rho =|\Psi\rangle \langle \Psi |$.  Consider a composite system consisting of two subsystems $A$ and $B$, with  $|a_i\rangle$ and $|b_j \rangle$, $i=1 \to N_a, j=1\to N_b$, as the orthogonal basis states in the subspaces of $A$ and $B$ respectively. An arbitrary  pure state $|\Psi \rangle$ of the composite system in the $N_a \times N_b$ product basis  $ |a_i\rangle \; |b_j \rangle$ can be written as

\begin{eqnarray}
|\Psi \rangle = \sum_{i,j}^{N_a, N_b}  C_{ij}  \; |a_i\rangle \; |b_j \rangle \label{p1}
\end{eqnarray}
Representation of  $\rho$ in the  product basis gives  pairwise products of the coefficients $C_{ij}$ as the elements of the density matrix $\rho$. A partial trace operation on $B$-subspace then leads to the reduced density matrix $\rho_A$ which turns out to be a Wishart matrix

\begin{eqnarray}
\rho_A  ={\rm Tr}_B \; \rho  \equiv C \cdot C^{\dagger}
\label{ws1}
\end{eqnarray}
subjected to a fixed trace constraint ${\rm Tr}\rho_A =1$. As $C$ consists of the coefficients (components) of the state $\Psi$, we refer it as the state matrix or just $C$-matrix for brevity.
The reduced density matrix $\rho_B$ can similarly be obtained by  projecting $\rho$ onto the orthogonal base of  $A$ respectively: $\rho_B ={\rm Tr}_A \; \rho$.

For the bipartite representation of a many body system, the entanglement measures between its two parts, can be obtained from the eigenvalues of the reduced density matrices  
of the subsystems. For example, the R\'enyi entanglement entropy $R_{\alpha}$ for $\rho_A$ is defined as 
\begin{eqnarray}
R_{\alpha} &=& {1\over 1-\alpha} {\rm log}_b \; {\rm Tr}\; (\rho_A)^{\alpha}
=  {1\over 1-\alpha} {\rm log}_b \; \sum_n \lambda_n^{\alpha}. 
\label{ra}
\end{eqnarray}
where $\lambda_n$ refers to the $n^{th}$ eigenvalue of the reduced density matrix $\rho_A$.
The limit $\alpha \to 1$ corresponds to the von Neumann entropy
\begin{eqnarray}
R_1 &=& - {\rm Tr}\; \rho_A \; {\rm log}_b \rho_A = - \sum_n \lambda_n \; {\rm log}_b \; \lambda_n 
\label{vn1}
\end{eqnarray}
with base $b$ referring to the dimension of the local Hilbert space of the subunits of the subsystems $A$ and $B$; (hereafter the base will be suppressed unless needed for clarity).
Further the min-entropy $R_{\infty}(\rho_A)$, defined as $\lim_{\alpha \to \infty} R_{\alpha} (\rho_A)$, can be expressed in terms of the maximum eigenvalue: $R_{\infty}(\rho_A) = - {\rm log}_b\; \lambda_{max}$.

\subsection{Structure of state matrix}

As the above definitions indicate, the determination of  entanglement entropies requires a prior knowledge of the reduced density matrices  and therefore the components of a pure bipartite state in the basis of interest. A physically motivated basis can often be chosen as the one that preserves the same global constraints (e.g. symmetry,  conservation laws etc.) as the system. The components of any state of the system, represented in such a basis, then satisfy the transformation rules for the constraint class. As a consequence, the structure   of the $C$ matrix, e.g., whether it is  real or complex,  the relation among the elements and their transformation rules etc., depends on the global  constraints e.g. symmetry,  conservation laws etc. For example, for systems with both time-reversal symmetry and integer angular momentum, a physically motivated basis can always be chosen as real. Thus $C$ is a real matrix for systems with both time-reversal symmetry and integer  angular momentum, a quaternion matrix for the systems with time-reversal symmetry and half integer angular momentum, and, a complex matrix in absence of the.time-reversal symmetry.
For clarity purposes, the present study is confined only to $C$ real or complex with no other matrix constraints. The elements of  $C$ can then be written as $C_{kl}=\sum_{s=1}^{\beta} (i)^{s-1}C_{kl;s}$ with $\beta=1$ or $2$ for $C$ real or complex; taking  $N_a=N, N_b=N+\nu_0$, we have $k=1 \to N, l=1 \to (N+\nu_0)$. (The generalization to quaternion $C$ can be done following similar steps but is technically tedious and is therefore not included here.) 

The main influence on the structure of $C$ matrix however comes from the nature of the  interactions between the subsystems A and B. For example, in ergodic limit or when almost all units of A are interacting with those of B with almost same average strength, the correlations between any  orthogonal base pair, with one of the pair belonging to  basis states of A and other to B, can be almost same for all such pairs. This in turn leads to almost all matrix elements of  $C$ of the same order. On the contrary, the presence of  interactions, local to subsytems only, are expected to cause  correlations varying from one orthogonal base pair to the other.  The strength of such correlations can also vary from the system to system e.g. decaying as power law or exponentially.  We note that contrary to a Hamiltonian matrix where a fully localized quantum dynamics leads to a diagonal matrix, the localization in this case affects the structure along the columns. For example, the $C$-matrix corresponding to a separable state has non-zero elements only along the first column. The decaying strength of matrix elements along columns therefore indicate a state with partial entanglement. Similarly a growth of entanglement reflects on the $C$-structure as increasing strength of the entries along the columns.   The parts (b-d) of figure 1 display the schematic pictures corresponding to three examples of partial entanglement along with a fully entangled state in part (a); we use these $C$-matrices later on for our numerical analysis too.

Based on the physical  as well as the local Hilbert space dimensions (i.e the number of states of a typical subunit in a subsystem), the number and strength of the components of a typical state in the product basis can vary. This in turn can lead to  different sparse structures, based on the  location of the ''cut'' or  ''partition'' i.e how the many body system is divided into two subsystems. Previous studies have often focussed on the evolution of the entanglement growth with changing location of the cut in time. Although the location also enters in our formulation, the present work is primary focussed on the evolution with changing system conditions while the cut is kept at a fixed location.

We note that the present analysis is not based on any real physical Hamiltonian and therefore role of the physical dimensions of the system or the subunits of each subsystem and the associated local Hilbert spaces is not directly obvious from the formulation. The theoretical results derived here and numerical examples considered are however applicable in general to any dimensionality and can be obtained by relevant changes. For example, if a subsystem X (=A or B) consists of $g_x$ subunits or particles, each associated with a local Hilbert spaces of dimensionality $b$, this implies $N_x = b^{g_x}$ and the logarithmic terms appearing in various entanglement measures are then defined on the base $b$.

\subsection{Multi-parametric distribution of coefficients: role of constraints}

As discussed above, the entanglement analysis of a quantum state requires a prior knowledge of its components in a basis of interest. It is however  technically non-trivial, almost impossible, usually, to determine the components exactly if the state is that of a many body system. A lack of detailed information about the complicated interactions usually manifests through a randomization of the components with large fluctuations at a local scale. Based on Berry's Gaussian random wave hypothesis in context of quantum chaotic systems \cite{berry}, a typical ergodic many body eigenstate constitutes a random Fock-space vector with independently  distributed Gaussian entries \cite{page}:  in terms of the $C$ matrix elements, this corresponds to the constraints on their first two moments only:  $\rm{\langle |C_{kl}|^2 \rangle ={1\over 2 \gamma}, \langle C_{kl} \rangle =0}$.  This in turn leads to a reduced density matrix for the subsystems described by the standard Wishart random matrix ensemble \cite{zhh, apsk}. Based on maximum entropy hypothesis, $C$-matrix then belongs to a maximum information entropy ensemble. But, as predicted by Page's law,  an ergodic state is also maximally entangled.  Thus an  ensemble of $C$ with maximum information entropy (Shannon) here corresponds to maximum entanglement entropy of the state too.

A typical many body quantum state need not be ergodic and not all its coefficients (i.e $C_{ij}$) are, in general, of the same order. 
Indeed the statistical behavior  of the $C$-elements depends on the local constraints too. Based on  the physical aspects of the system, these can be of various types. While the global constraints (matrix constraints) affect nature of each matrix, the local constraints (ensemble constraints) influence distribution parameters of the matrix elements: disorder, dimensionality, boundary conditions, nature of dynamics (chaotic or non-chaotic), interactions among subunits of system etc. For example, assuming all independent elements, a  most generic form of $\rho_c$ can be written as 

\begin{eqnarray}
\rho_c(C) =  \mathcal{N} \; \rho_c\left(C_{11}, C_{12}, \ldots, C_{NN}\right)  \; F_c
\end{eqnarray}
with $F_c$ dependent on the matrix constraints including  ${\rm Tr} (C^{\dagger} \cdot C) =1$. The corresponding ensemble of $N \times N$ reduced density matrices $\rho_A = C \cdot C^{\dagger}$ then forms a generalized Wishart ensemble. With known constraints on  the $1^{\rm st}$ and  $2^{\rm nd}$ order moments of $C_{kl}$, a typical quantum state can be described by a Gaussian ensemble of $C$ matrices. The  ensemble constraints on  higher order moments of $C_{kl}$ in general  lead to a non-Gaussian ensembles.

A reduced density matrix for a typical pure quantum state of a complex system in general belongs to a  non-stationary Wishart ensemble with arbitrarily randomly distributed elements of $C$, with ensemble parameters governed by correlations among the basis states. The change in complexity e.g. many body interactions within the system can cause a variation of correlations between subsystems and thereby leading to evolution of the $C$-matrix elements. This is discussed in next section.

\subsection{Diffusion of matrix elements: ensemble complexity parameter}

A typical element of $C$-matrix describes the correlations between the two orthogonal bases with its distribution  a measure of the inaccuracy (deterministic or due to disorder) arising due to complexity of the system. The changing system conditions over time
can however lead to a variation of these correlations as well as the inaccuracy and thereby distribution parameters. As a consequence, the ensemble density $\rho_c$ can evolve due to change of the matrix elements as well as the parameters.          
As discussed below, the evolution of the ensemble density  in matrix space can be mimicked by that in the parameter-space; this in turn helps better insights in the behavior of entanglement measures.

As our analysis is based on many types of parameters, it is important to distinguish them at the onset: hereafter the term ''system parameters" refer to those which appear in a single matrix due to their presence in the corresponding operator or the chosen basis, the term ``distribution parameters"  or  ``ensemble parameters'' refers to those characterizing an ensemble of the matrices. Another set of parameters, referred  as the complexity parameters and an important aspect of our analysis,  is introduced later.

For clear exposition of our ideas, we consider the elements of $C$ as  independent Gaussian distributed, with  arbitrary mean and variances: 

\begin{eqnarray}
{\rho_c}(C; h,b)    &=&  \mathcal{N} \;  {\rm exp}\left[-  \sum_{k,l,s} \;{1\over 2 h_{kl;s}} \left( C_{kl;s} -b_{kl;s} \right)^2 \right] 
\label{rhoc}
\end{eqnarray}
with $\sum_{k,l,s}  \equiv \sum_{k=1}^N \sum_{l=1}^{N+\nu_0} \sum_{s=1}^{\beta}$ and  $\mathcal{N} $ as a normalization constant: $\mathcal{N} = \prod_{k,l, s} (2 \pi h_{kl;s})^{-1/2} $.  Here $h \equiv \left[h_{kl,s} \right]$ and $b\equiv \left[b_{kl,s} \right]$ refer to the matrices of variances and mean values of $C_{kl;s}$. Clearly, with different choices of $h$ and $b$-matrices, eq.(\ref{rhoc}) can give rise to many coefficient ensembles; some of them are used later in section IV for numerical verification of our results.

A small perturbation of the quantum  state by 
change of the system parameters over time causes the matrix elements $C_{kl}$ undergo a dynamics, thereby  leading to an evolution of $\rho_c$ in the matrix space. But as $h, b$ depend on system parameters too (i.e perturbation also affects the accuracy of each $C_{kl}$),  $\rho_c$ also evolves in the ensemble space, due to change in the parameters $h_{kl;s} \rightarrow h_{kl;s}+\delta h_{kl;s}$ and $b_{kl} \rightarrow b_{kl}+\delta b_{kl}$  over time. As discussed in   {\it appendix A} (also in \cite{ptche} in context of chiral Hamiltonians), the multi-parametric evolution of $ \rho_c(C)$,  generated  by a combination of first order parametric derivatives $T\equiv \sum_{k\le l;s}\left[(1-\gamma h_{kl;s}) \; {\partial \over\partial h_{kl;s}} - \gamma b_{kl;s} {\partial \over\partial b_{kl;s}}\right]$,   leads to a  diffusion with finite drift  in $C$-matrix space. It is however possible to define a transformation of ensemble parameters from $h, b$-set to another set  of $M$ variables, namely, $y_1, y_2,\ldots y_M$ that leads to $T \rho \equiv  \frac{\partial \rho}{\partial y_1}$. Thus the transformation reduces  the multi-parametric evolution  of the ensemble density to a single parametric diffusion, with only $Y \equiv y_1$ varying and rest $y_k, k >2$ as the constants of evolution. Using $\rho = \mathcal{N}_0 \; \rho_c$, we have 

\begin{eqnarray}
\frac{\partial \rho}{\partial Y} =  \sum_{k,l,s} \frac{\partial}{\partial C_{kl;s}}\left[ \frac{\partial  \rho}{\partial C_{kl;s}}+\gamma \; C_{kl;s} \; \rho \right]
\label{ch8}
\end{eqnarray}
with  ${\mathcal N}_0 ={\rm e}^{\int f_0(Y)\; {\rm d}Y}$ and $f_0(Y) = T \log {\mathcal N}$.
Here the evolution parameter  $Y = Y(\{h_{kl;s}, b_{kl;s}\}$ is a functional: it is a function  of all the distribution parameters which in turn are the  functions of system parameters and the latter  in turn are functions of time. As it contains information about the complexity of the ensemble,   $Y$ is known as the ensemble complexity parameter

\begin{eqnarray}
Y=-\frac{1}{2 M\gamma} \; {\rm ln} \left[\prod_{k,l}' \prod_{s=1}^{\beta} |(1- 2 \gamma h_{kl;s})  | \;|b_{kl;s}|^2\right]+{\rm const}
\label{y}
\end{eqnarray}
with $\prod_{k,l}'$  implies a product over non-zero $b_{kl;s}$ as well as $x_{kl;s}=(1-\gamma h_{kl;s})$, with $M$ as their total number; (for example for the  case with all  $x_{kl;s} \not=0$ but $b_{kl;s}=0$, we have $M=\beta N (N+\nu_0)$ and for case with all  $x_{kl;s} \not=0$ and $b_{kl;s} \not=0$, we have $M=2\beta N (N+\nu_0)$). Further  $\gamma$ is an arbitrary parameter, related to final state of the ensemble  (giving the variance of  matrix elements at the end of the evolution) and  the constant in eq.(\ref{y}) is determined by the initial state of the ensemble. 

Eq.(\ref{ch8}) describes the evolution of $\rho$, equivalently, $\rho(C; Y)$ at $Y =y_1(h,b)$ starting from an arbitrary initial ensemble density, say $\rho_0(C,Y_0)$ at $Y_0=y_1(h_0, b_0)$ with $h_0, b_0$ as the initial ensemble parameters. As discussed in {\it appendix} A, (following from eq.(\ref{hby})), the evolution is subjected to conditions $y_k={\rm constant}$ for $k >1$ and reaches a steady state when ${\partial \rho/\partial Y}\rightarrow 0$ with $\rho$ approaching the probability density of the stationary Wishart ensemble, $\rho \propto {\rm e}^{-(\gamma/2) {\rm Tr}C^2}$. As the system information in eq.(\ref{ch8}) enters only through $Y$, its solution $\rho(C; Y)$  remains same for different ensembles, irrespective of the details of their $h$ and $b$ matrices,  if they share same values for $y_1, y_2, \ldots, y_M$.

\subsection{Diffusion of Schmidt eigenvalues}

With $\rho_A$ as a $N \times N$ Hermitian matrix, its eigenvalue equation can be written as $\rho_A \; U = U \; \Lambda$ where
 $U$ is  the $N\times N$ eigenvector matrix, unitary in nature i.e $U^{\dagger}  U=1$ ($\rho_{A}$ being  Hermitian) and $\Lambda$ is the the $N\times N$ diagonal matrix of  its eigenvalues,  $\Lambda_{mn}= \lambda_n \; \delta_{mn}$ also referred as the Schmidt eigenvalues. A small change $\delta Y$ in $Y$ (given by eq.(\ref{y}) results in 
 diffusive dynamics of of $C$ that in turn manifests in the $\rho_A$-matrix space too and the moments for  the matrix elements $\rho_{A;mn}= \sum_{k=1}^{N} C_{mk} C^*_{nk}$ can be calculated from those  of $C$. 

To proceed further, it is important to note that eq.(\ref{ch8}) is equivalent to the evolution of an arbitrary $N_a \times N$  (with $N_a \ge N$) rectangular matrix $C_0$ subjected to a random perturbation, of strength $t$, by another $N_a \times N$  rectangular matrix $V$ (real or complex for $L$ real-symmetric or complex Hermitian, respectively,  with  Gaussian density $\rho_v(V) =\left(\frac{1}{2 \pi v^2}\right)^{\beta N_a N/2} {\rm e}^{-{1\over 2 \; v^2} \; {\rm Tr} (V V^{\dagger})}$).
The perturbed matrix $C(t)$ is described  as $C(t)=\sqrt{f} (C_0+t \; V)$ with $f=(1+ \gamma_0 t^2)^{-1}$, $C(0)=C_0$ as a fixed random matrix and $\gamma_0$ as an arbitrary positive constant (\cite{sp}).  The evolution due to variation of strength $t$ of the random perturbation  is Markovian  if considered in terms of a rescaled parameter $Y=-{1\over 2 \gamma} \;  \ln f ={1\over 2 \gamma} \; \ln (1+ \gamma \; t^2)$ \cite{sp, pslg}:

\begin{eqnarray}
C(Y)  &\equiv&  C(0) \; {\rm e}^{-\gamma Y} +  V(Y) \; \left({1- {\rm e}^{-\gamma Y} \over \gamma}\right)^{1/2}
 \label{att0} 
\end{eqnarray} 
As discussed in {\it appendix B},  relevant information for the moments of $\lambda_n$ can subsequently be derived by using second order perturbation theory for Hermitian matrices which on substitution in standard Fokker Planck equation leads to the    equation describing $Y$-governed evolution of the joint probability distribution function ({\it jpdf})  $P_{\lambda}(\lambda_1, \lambda_2, \ldots,  \lambda_N; Y)$,
 
\begin{eqnarray}
\frac{\partial P_{\lambda}}{\partial Y}= \sum_{n=1}^N \left[\frac{\partial^2 (\lambda_n \; P_{\lambda})}{\partial \lambda_n^2} - \frac{\partial}{\partial \lambda_n}\left( \sum_{m=1}^N \frac{\beta \lambda_n}{\lambda_n- \lambda_m} - {\beta \nu} - 
 2 \gamma \lambda_n  \right)  P_{\lambda}\right]
\label{pdl1}
\end{eqnarray}
where $\nu=(N_b-N_a-1)/2 =(\nu_0-1)/2$. (Here we set $v^2=1/4$ for simplification and without loss of generality). Hereafter unless required for clarity, we will use the notation $f(\lambda)$ for $f(\lambda_1, \ldots, \lambda_N)$ for any arbitrary function $f$.

The above equation describes the diffusion of $P_{\lambda}(\lambda,Y)$, with a finite drift, from an arbitrary initial state $P_{\lambda}(\lambda_0,Y_0)$ at $Y=Y_0$. In limit $\frac{\partial P_{\lambda}}{\partial Y} \rightarrow 0$ or $Y\rightarrow \infty$, the diffusion approaches a unique steady state: 

\begin{eqnarray}
P_{\lambda}(\lambda; \infty) = C_{\beta} \; \prod_{m < n=1}^N |\lambda_m -\lambda_n|^{\beta} \; \prod_{k=1}^N |\lambda_k|^{2 \nu \beta -1} \;  {\rm e}^{-{\gamma \over 2 } \sum_{k=1}^N \lambda_k}
\label{psch}
\end{eqnarray}
with $C_{\beta}$ as the normalization constant. The above {\it jpdf} of the eigenvalues corresponds to that of a stationary Wishart ensemble and is  consistent with the expectation: the  limit  ${\partial \rho\over  \partial Y} \to 0$ of eq.(\ref{ch8}) corresponds to   ${\partial P_{\lambda} \over  \partial Y} \to 0$ for eq.(\ref{pdl1}), and, with $\rho$  approaching   stationary Wishart ensemble density in this limit,  $P_{\lambda}$ is expected to approach corresponding {\it jpdf} of the eigenvalues.  We note that eq.(\ref{pdl1}) can also be derived by an exact diagonalization of eq.(\ref{ch8}) (following the similar route as discussed in \cite{ptche} for the chiral ensembles).

As mentioned in section II.C, the reduced density matrix $\rho_A$ is a Wishart matrix subjected to a fixed trace constraint ${\rm Tr}\rho_A =1$.  The {\it jpdf}  $P_c(\lambda_1, \lambda_2, \ldots,  \lambda_N; Y)$ (denoted as $P_{c}(\lambda; Y)$ hereafter)  as of the eigenvalues of $\rho_A$ can then be given as,
\begin{eqnarray}
P_{c}(\lambda; Y) =C_{hs} \; \delta(\sum_n \lambda_n -1) \; P_{\lambda}(\lambda, Y).
\label{pc} 
\end{eqnarray}
with $C_{hs} \equiv C_{hs}(Y)$ is the normalization constant for $P_{c}(\lambda; Y)$; we note here that the constraint $\sum_n \lambda_n =1$ confines  $P_{c}(\lambda; Y)$ to an area much less than that of $P_{\lambda}(\lambda, Y)$. Further, the stationary Wishart ensemble subjected to the above constraint is also referred as the Hilbert-Schmidt ensemble.

\section{Growth of Average quantum information entropies}

Following from the definition (\ref{ra}), the average entanglement measure, say $R_f \equiv f(\lambda)$ for a typical state can be expressed in terms of the {\it jpdf} $P_c$ of the Schmidt eigenvalues

\begin{eqnarray}
\langle f \rangle= C_{hs} \; \int f(\lambda) \; P_{\lambda}(\lambda) \; \delta(\sum \lambda_n-1)  {\rm D} \lambda.
\label{rr}
\end{eqnarray}
For simplification of the technical analysis, the above can be rewritten as $\langle f(S_1=1)\rangle$ where 

\begin{eqnarray}
\langle f (S_1) \rangle= C_{hs} \; \int  \;  f(\lambda) \; \; P_{\lambda}(\lambda)  \; \delta(S_1-\sum \lambda_n) \;  {\rm D} \lambda
\label{rr1}
\end{eqnarray}
The above relation can now be used to derive the $Y$-governed evolution equation of $\langle f \rangle$: 

\begin{eqnarray}
\frac{\partial  \langle f (S_1) \rangle }{\partial Y}=   \int   \;  f(\lambda) \; \frac{\partial (C_{hs} \; P_{\lambda} )}{\partial Y}  \; \delta(S_1-\sum \lambda_n) \; {\rm D} \lambda
\label{rr2}
\end{eqnarray}
Substitution of eq.(\ref{pdl1})  then leads to 

\begin{eqnarray}
\frac{\partial  \langle f (S_1) \rangle }{\partial Y}= I_0 + I_1 + I_2
\label{rr3}
\end{eqnarray}
with $\delta_1 \equiv \delta(S_1-\sum \lambda_n)$, $I_0 = \frac{\partial \log C_{hs}}{\partial Y}  \langle f (S_1) \rangle$  and

\begin{eqnarray}
I_1 &=& - \; C_{hs} \; \sum_{n=1}^N \int  {\rm D} \lambda \;  \delta_1 \; f(\lambda)   \frac{\partial}{\partial \lambda_n}\left( \sum_{m=1}^N \frac{\beta \lambda_n}{\lambda_n- \lambda_m} - {\beta \nu} - 2 \gamma \lambda_n  \right)  P_{\lambda}
\label{i1}
\end{eqnarray}
and,
\begin{eqnarray}
I_2 &=& C_{hs} \; \sum_{n=1}^N \int  {\rm D} \lambda \;  \delta_1 \; f(\lambda) \;  \frac{\partial^2 (\lambda_n \; P_{\lambda})}{\partial \lambda_n^2}  \label{i2}
\end{eqnarray}
As discussed in {\it appendices} C and D, using integration by parts along with $\frac{\partial \delta_1}{\partial \lambda_n}  = - \frac{\partial \delta_1}{\partial S_1}$, $I_1$ and $I_2$ can further be simplified leading to a close form equation for $\langle f \rangle$ which can be solved, in principle, to obtain the average entanglement measures for any arbitrary $Y$ value.
Below we consider the cases  for $f(\lambda)=R_1(\lambda)$ and $R_2(\lambda)$.

\subsection{Average Von-Neumann Entropy}

 The ensemble average of von Neumann entropy $R_1 = - \sum_n \lambda_n \log \lambda_n$, defined in eq.(\ref{vn1}),  is given by $\langle R_1(1) \rangle$ where, from eq.(\ref{rr}), we have
 
\begin{eqnarray}
\langle R_1(S_1) \rangle= C_{hs} \; \int  \left[ - \sum_n \lambda_n \log \lambda_n \right] \; P_{\lambda}(\lambda) \; \delta(S_1-\sum \lambda_n)  \; {\rm D} \lambda.
\label{rvn1}
\end{eqnarray}
with $S_1=\sum_n \lambda_n$.

Using $f= - \sum_n \lambda_n \log \lambda_n$ in eq.(\ref{rr1}) and substituting $I_1$ and $I_2$ from {\it appendices} C and D, the $Y$-governed growth of average von Neumann entropy, from an arbitrary initial state at $Y=Y_0$, can be given as,
\begin{eqnarray}
\frac{\partial \langle R_1(S_1) \rangle}{\partial Y} &=& 
\alpha  \; \langle R_1 \rangle +  {{1\over 2}  \beta N_{\nu}} \; \langle R_0 \rangle  - {1\over 2} (\beta N N_{\nu} +\beta N(N-1)- 4 \; \gamma \; S_1 +2 (N-2)) J+G_s \nonumber 
\\\label{rvn2}
\end{eqnarray}
with  $\alpha = {\partial \log C_{hs}\over \partial Y}$, $N_{\nu}=N - 2 \nu-1$, and 
\begin{eqnarray}
G_s(S_1) &=& -{1\over 2} \left( \beta N N_{\nu} - 4 \gamma S_1 \right) \; \frac{\partial  \langle R_1\rangle }{\partial S_1} 
 + S_1 \; \frac{\partial^2 \langle R_1 \rangle }{\partial S_1^2}  + 2 \frac{\partial J }{\partial S_1}
\label{gs1} \\
\langle R_0 \rangle &=& -\langle \sum_n \log \lambda_n \rangle, \label{r0} \\
J \equiv J(S_1) &=&  C_{hs} \; \int  {\rm D} \lambda \;  \delta_1 \; P_{\lambda},
\label{j} 
\end{eqnarray}

{\bf Large $N$ case:} 
Noting  that $Y$ is inversely proportional to $M = N\,N_{\nu}$, a rescaling of $Y$ in eq.(\ref{rvn2}) can  reduce its technical complexity; the resulting differential equation using the rescaled parameter $\Lambda = N\, N_{\nu} \, (Y-Y_0)$ is
  \begin{equation}
    {2\over \beta}  \frac{\partial \langle R_1 \rangle}{\partial \Lambda} =\left( {\langle R_0 \rangle\over N}- q_0 \; J\right) - \frac{\partial \langle R_1 \rangle}{\partial S_1},
    \label{vnd1}
  \end{equation}
with $q_0 =\frac{N_{\nu}-(N-1)}{N_{\nu}}$.

The above equation can be solved by using the standard method  of characteristics for partial differnetial equations. Noting that a substitution of $\gamma=0, g(\Lambda, S_1)={\langle R_0 \rangle\over N}- J$  reduces eq.(\ref{df1}) to eq.(\ref{vnd1}),  its general  solution can be given as 
(details discussed in {\it appendix F}),
  \begin{equation}
    \langle R_1 (S_1, \Lambda ) \rangle = B + \int  {dS_1} \left({\langle R_0 \rangle\over N}- q_0 \; J\right) - \exp\left(\tau \; S_1 -\frac{\tau \,\beta \,\Lambda }{2}\right)
    \label{r1s2}
  \end{equation}
with constants $B$ and $\tau$  to be determined from the boundary conditions. As $\langle R_1 (S_1, 0) \rangle=0$  for a separable initial state chosen at $Y = Y_0$ i.e $\Lambda=0$, we have 
\begin{eqnarray}
{e}^{\tau S_1} =I_1 = B + \int \left({\langle R_0 \rangle\over N}-q_0 \;  J\right) \; {dS_1}.
\label{cc}
\end{eqnarray}
 This on substitution in eq.(\ref{r1s2}) leads to

  \begin{equation}
    \langle R_1 (S_1, \Lambda ) \rangle =  I_1 \;  \left(1- \exp\left(-\frac{\tau \,\beta \,\Lambda }{2}\right)\right)
    \label{r1s3}
  \end{equation}

Further, to satisfy the boundary condition  that $\langle R_1 (S_1, \Lambda ) \rangle $ has an upper bound $ R_{\infty}$ for $\Lambda \to \infty$ i.e  we must have $R_{1,\infty} = I_1$. The latter then gives, for arbitrary $S_1$, 
  \begin{equation}
    \langle R_1 (S_1, \Lambda ) \rangle =R_{1,\infty} \, \left[1 - \exp\left(-\frac{\tau \,\beta \,\Lambda}{2} \right)\right]    \label{r1s4}
  \end{equation}
 with $\tau$  determined by eq.(\ref{cc}).

To determine $\tau$ from eq.(\ref{cc}), we note  that $\langle R_0 \rangle$  changes rapidly with $\Lambda$ (relative to  $\langle R_1 \rangle$ and with $R_0$ defined in eq.(\ref{r0})) and can be approximated by its equilibrium value at $\Lambda \to \infty$ (corroborated by the numerical analysis of three ensembles discussed in section IV). Further $J(S_1)$ given by eq.(\ref{j}) is a normalization constant for $P_{\lambda}(\lambda_1, \ldots, \lambda_N; Y)$ subjected to trace constraint $S_1=\sum_N \lambda_n$ and is kept fixed as $Y$ and thereby $\Lambda$  evolves.
For $S_1=1$, this leads to $\tau = \log \left(B+ \left({\langle R_0 \rangle\over N}-q_0 \, J\right) \right) = \log R_{1,\infty}$.

 For small $\Lambda$, the above equation then gives a linear increase of the entropy with $\Lambda $: $\langle R_1 (\Lambda )\rangle \approx   {1\over 2} \; \beta \;  \left(R_{1,\infty} \; \log R_{1,\infty} \right)\; \Lambda $. For large $\Lambda $, the growth  is again rapid, with $\langle R_1 (\Lambda )\rangle$ approaching the constant value  $R_{1,\infty} \approx   \left( {\langle R_0 \rangle\over N}- q_0 \, J \right)+ B$. By substituting  large $\Lambda$ result for $\langle R_0 \rangle \approx N \log N + N $ (i.e stationary Wishart limit as derived in \textit{appendix H})), using normalization $J(1)=1$ and setting the arbitrary constant $B=0$, we then arrive as  the expected limit for the stationary states $\langle R_1 (\infty)\rangle \sim \left(\log N - \frac{N}{N_{\nu}}\right)$. In addition, this is also consistent with the expected Page's limit \cite{page} up to a factor of $2$.

As eq.(\ref{r1s4}) indicates, the growth of $R_1(\Lambda)$ with changing complexity  is indeed sensitive to the underlying symmetry conditions (through parameter $\beta$) as well as to  the ''size cut'' of subsystems through parameter $\nu = (\nu_0-1)/2=(N_b-N_a-1)/2$; the latter enters in the formulation through $\langle R_0\rangle$ as well as $\Lambda= N N_{\nu} (Y-Y_0)$.  We note that, for system conditions changing with time, $Y$ and thereby $\Lambda$ are indeed  a function of time; the above analysis then describes the time-dependent growth too (discussed later in section III.c).

\vspace{0.2in}

{\bf Finite $N$ case:} Due to rescaling and subsequently neglecting the terms in eq.(\ref{rvn2}),  the solution (\ref{r1s4}) is valid only in large $N$ limit. As  the local Hilbert space of a subsystem can also be of finite size e.g. a qubit in contact with other qubits, it is relevant to consider finite $N$ case too.

To solve eq.(\ref{rvn2}) for finite $N$ case, we note that $R_1$ and $S_1$ vary at different rates with $Y$ as well as $N$. This permits us to  assume  $\langle R_1 (S_1, Y) \rangle = g(S_1) \; \langle  R_1(1) \rangle$ (based on multiplication of independent probabilities) with $\langle R_1(1) \rangle \equiv \langle R_1(S_1=1,Y) \rangle$ and $g(S_1)$ as a function of $S_1$ only (with its exact form not needed  for our qualitative analysis). Similarly we can write $J(S_1)=g_0(S_1) \; J(1)$. 
 This in turn gives $G_s(S_1)=\phi(S_1) \; \langle R_1(1) \rangle$ with
 
\begin{eqnarray}
\phi(S_1) = -{1\over 2} \left( \beta N N_{\nu} -4 \gamma \; S_1\right) \; g'(S_1) 
+ S_1 \; g''(S_1) + 2 g'_0(S_1)
\label{gs2a}
\end{eqnarray}
with $g'(S_1) \equiv \frac{\partial g(S_1)}{\partial S_1}$, $g''(S_1) \equiv \frac{\partial^2 g(S_1)}{\partial S_1^2}$ and $g'_0(S_1) =\frac{\partial g_0}{\partial S_1}$. (Alternatively we can also first expand $\langle R_1 (S_1) \rangle$ in a Taylor's series near $S_1=1$: $\langle R_1 (S_1) \rangle= \langle R_1 (1) \rangle + \sum_k {\partial_s^k  \langle R_1\rangle } \mid_{S_1=1} {(S_1-1)^k \over k!} $ (with notation $\partial_s \equiv  \frac{\partial }{\partial S_1}$) and use the assumption that ${\partial_s \langle R_1(S_1)\rangle }= g'(S_1) \; \langle R_1(1) \rangle$. This  leaves the final result unaffected). For later reference, we note that $g'(1) >0$ (numerical verification depicted in figure $6$); this can be seen directly by considering  ${\delta R_1 \over \delta S_1}$.)

As our interest here is in  case $S_1=1$ only, eq.(\ref{rvn2}) can be rewritten as (using notation $\langle R_1\rangle \equiv \langle R_1(1) \rangle$ and $J(1)=1$),

\begin{eqnarray}
\frac{\partial \langle R_1 \rangle}{\partial Y}= v_2 -v_1 \langle R_1 \rangle
\label{rvn2a}
\end{eqnarray}
with $v_1(Y)=-(\alpha - 4 \gamma) g(S_1) -\phi(1)$ and $v_2(Y)={{1\over 2}  \beta N_{\nu} } \; \langle R_0 \rangle  + {1\over 2} (\beta  N N_{\nu}   +2 (N-1) - 4 \; \gamma ) $

The solution of the above equation for arbitrary initial condition and for finite $Y$ is

\begin{eqnarray}
\langle R_1 (Y)\rangle = {\rm e}^{-\int_{Y_0}^Y   {\rm d}Y \;   v_1(y) }\; \left[\int_{Y_0}^Y {\rm d}t \; v_2(t) {\rm e}^{\int_{Y_0}^t v_1(t) \; {\rm d}t} +R_1(Y_0) \right]
\label{r1sa}
\end{eqnarray}

As the finite $N$ solution discussed above is based on some approximations, it is relevant to analyze its large $N$ limit and check its consistency with large $N$ solution discussed above (obtained by rescaling of $Y$ and without making any approximation). As discussed in {\it appendix G}, the two solutions are indeed consistent in large $N$-limit; this in turn supports the approximations made above to simplify finite $N$ cases.

It is worth noting here the important role played by the terms originating from the derivative ${\partial R_1(S_1) \over \partial S_1}\mid_{S_1=1}$ i.e the terms which reflects the role of the constraint $S_1=\sum_n \lambda_n$. These terms are necessary to reach the correct stationary limit of $\langle R_1\rangle $  i.e the one for the Hilbert-Schmidt ensemble and thereby indicate the role played by the collective dynamics of eigenvalues. In the unconstrained  stationary limit i.e stationary Wishart ensemble, $\lambda_n$ are free to move on positive real axis although the Gaussain term in eq.(\ref{psch})  for $P_{\lambda}(\lambda)$ confines them from escaping to infinity; the competition between repulsion term and confinement due to Gaussian term results in on an average $\lambda_n \sim {1\over \sqrt{N}}$ and thereby  $R_1(\lambda) \sim {1\over 2} \sqrt{N} \log N$. With $R_1 \sim \log N$ for the constrained case, clearly the von-Neumann entropy is reduced due to the  constraint.

\subsection{Average 2nd order Renyi Entropy}

As in the case of $R_1$,  the ensemble average of 2nd Renyi entropy $R_2$, 
defined in eq.(\ref{vn1}),  is given by 
$\langle R_2(S_1=1) \rangle$ where, from eq.(\ref{rr}), we have

\begin{eqnarray}
\langle R_2(S_2) \rangle= C_{hs} \; \int \; \delta(S_1-\sum \lambda_n) \; \left[-\log \sum_n \lambda_n^2 \right] \; P_c(\lambda) \; {\rm D} \lambda
\label{rc2}
\end{eqnarray}

Using $f= R_2$ in eq.(\ref{rr1}) and substituting $I_1$ and $I_2$ from {\it appendices} C and D, 
the $Y$-governed growth of average 2nd order Renyi entropy can be given as 

\begin{eqnarray}
\frac{\partial \langle R_2(S_1)\rangle}{\partial Y}= 
(\alpha + 2\gamma) \langle R_2 \rangle -\left(2\beta (N-\nu-1)+2\right) \;S_1 \; \langle {1\over S_2} \rangle + 4 \langle {S_3 \over S_2^2} \rangle  + 4 \; \gamma \; J  +G_s
\label{r2f}
\end{eqnarray}
with

\begin{eqnarray}
G_s(S_1) = 4 \; \frac{\partial J }{\partial S_1}  + \left({2 \gamma S_1} + 2 -(1/2) \beta N N_{\nu} \right) \; \frac{\partial  \langle R_2 \rangle }{\partial S_1}+ S_1 \; \frac{\partial^2 \langle R_2 \rangle }{\partial S_1^2}  
\label{gs2}
\end{eqnarray}

{\bf Large $N$ case:} proceeding as in $\langle R_1\rangle$ case, we again consider  a rescaling of $Y$; eq.(\ref{r2f}) in terms of the rescaled parameter $\Lambda=N\, N_{\nu} \, (Y-Y_0)$ can be written as


\begin{equation}
    {2\over \beta}  \frac{\partial \langle R_2 \rangle}{\partial \Lambda} =-\eta \, S_1  \, \langle {1\over S_2} \rangle  - \frac{\partial \langle R_2 \rangle}{\partial S_1},
    \label{rn2}
  \end{equation}
  with $\eta = {4 (N-\nu-1)\over N N_{\nu}} $. Again noting that a substitution of $\gamma=1, g(\Lambda, S_1)=-\eta \, \langle {1\over S_2} \rangle  $  reduces eq.(\ref{df1}) to eq.(\ref{rn2}),  its general  solution can be given as (details discussed in {\it appendix F}),
  \begin{equation}
    \langle R_2 (S_1, \Lambda ) \rangle =  B-\eta \; \int  {dS_1} \; \langle {1\over S_2} \rangle \; S_1- \exp\left(\tau \; S_1 -\frac{\tau \,\beta \,\Lambda }{2}\right).
 \label{rn2a}
  \end{equation}
Here both $B$ and $\tau$ are unknown constants and can be determmined by imposing the boundary conditions at $\Lambda=0$ and $\infty$. Again using the separable initial state at  $\Lambda=0$, we have 
\begin{eqnarray}
{e}^{\tau S_1}  =I_2 \equiv  B- \eta \int \; {dS_1} \; \langle {1\over S_2} \rangle \; S_1 
\label{cc1}
\end{eqnarray}
 This on substitution in eq.(\ref{rn2a}) leads to $ \langle R_2 (S_1, \Lambda ) \rangle =  I_2 \;  \left(1- \exp\left(-\frac{\tau \,\beta \,\Lambda }{2}\right)\right)$. Applying the boundary condition  that $\langle R_2 (S_1, \Lambda ) \rangle $ has an upper bound $ R_{2,\infty}$ for $\Lambda \to \infty$ now gives $R_{2,\infty} = I_2$.  For arbitrary $S_1$, we now have
  \begin{equation}
    \langle R_2 (S_1, \Lambda ) \rangle =R_{2,\infty} \, \left[1 - \exp\left(-\frac{\tau \,\beta \,\Lambda}{2} \right)\right]    \label{rn2b}
  \end{equation}
 with $\tau$  now given by eq.(\ref{cc1}) and can be  calculated if $S_1$-dependence of $\langle {1\over S_2} \rangle$ is known.

As in case of $R_1$, eq.(\ref{rn2b}) gives, for small $\Lambda$, a linear increase  with $\Lambda$, 
\begin{eqnarray}
\langle R_2 (\Lambda)\rangle \approx   \left({1\over 2} \, \beta \; \tau \, R_{2,\infty}\right)   \,  \Lambda .
\label{r2sy}
\end{eqnarray}
We note however that $\tau$ is now different (from that of $R_1$ case ) and is given by eq.(\ref{cc1}). For large $\Lambda$, the growth is again exponentially rapid, with $\langle R_2 (\Lambda)\rangle$ approaching a constant value $R_{2,\infty} = I_2$.  To calculate the  integral $I_2$ explicitly  we  need information about  $S_1$-dependence of $\langle {1\over S_2} \rangle$. Based on our numerical analysis of three different ensembles (discussed in section IV and displayed in figure $5$), we  find 
\begin{eqnarray}
\langle {1\over S_2} \rangle  \sim q_0 \; (1- {\rm e}^{-q \, \Lambda)})
\label{s2}
\end{eqnarray}
 with constant  $q$ dependent on the ensemble and $q_0 =\lim_{\Lambda \to \infty} \langle {1\over S_2} \rangle \sim \alpha N \log N$. A substitution of $\Lambda ={\beta S_1\over 2} +C_0$ (as discussed in {\it appendix F}) in eq.(\ref{s2}), subsequently using it in eq.(\ref{cc1})  then gives $I_2 =B-\eta \; q_0/2 $   and thereby $\tau \approx \log(B- \eta \; q_0/2 )$ for $S_1=1$. Choosing the arbitrary constant $B$ now appropriately i.e $B  \sim (4 \alpha +1) \log N$, we now achieve correct stationary limit $R_2 \sim \log N$ as  $\Lambda \to \infty$.

Eq.(\ref{rn2b}) describes  the growth of $\langle R_2(\Lambda)\rangle $ with changing complexity: similar to $R_1$, $R_2$  also depends, for finite $\Lambda$ cases, on the symmetry parameter $\beta$ as well as  the ''size cut''  $\nu$. Another important result derive by the above analysis is as follows: while both $\langle R_1(\Lambda)\rangle $  as well as $\langle R_2(\Lambda)\rangle $ show a linear increase of  for small $\Lambda$, the rate of increase $\tau$ is different for them (as given by eq.(\ref{cc}) and eq.(\ref{cc1}) for the two cases). The different rate of increase for the two average entropies is also corroborated by our numerical analyis discussed in section IV.

\vspace{0.2in}

{\bf Finite $N$ case:} The solution (\ref{rn2b}) of eq.(\ref{r2f}) is valid only in large $N$ limit. To determine the solution  for finite $N$ case, we again proceed as in $\langle R_1 \rangle$ case and  rewrite $G_s(S_1) = \phi(S_1) \; \langle R_2(1) \rangle$ with

\begin{eqnarray}
\phi(S_1) \equiv S_1 \; g'_0(S_1)+ 4 \left({2 \gamma S_1} + 2 -(1/2) \beta N N_{\nu}\right) \; g'(S_1) + S_1 \; g''(S_1),
\label{phi}
\end{eqnarray}
We note that, contrary to $R_1$ case, now  $g'(1) < 0$ (numerical verification displayed in figure $6$). (The latter can also be seen directly by writing $R_2$ as $R_2= - \log S_2$, $S_q= \sum_n^{N} \lambda_n^q$  which gives   
${\delta R_2 \over \delta S_1} = -{1\over S_2} \; {\delta S_2 \over \delta S_1} \approx -{2\over S_2} <0$).  The above on substitution in eq.(\ref{r2f}) gives

\begin{eqnarray}
\frac{\partial \langle R_2 \rangle}{\partial Y}= u_2 -u_1 \langle R_1 \rangle
\label{rn2aa}
\end{eqnarray}
with $u_1(Y)=-(\alpha + 2\gamma) \; g(S_1) - \phi(1)$ and $u_2(Y)=-\left(2\beta (N-{\nu}-1)+2\right) \; \langle {1\over S_2} \rangle + 4 \langle {S_3 \over S_2^2} \rangle  + 4 \; \gamma +1  $, the general solution for $\langle R_2 \rangle$ for arbitrary initial condition and for finite $Y$ can be written as 

\begin{eqnarray}
\langle R_2 (Y)\rangle = {\rm e}^{-\int_{Y_0}^Y u_1(Y) \; {\rm d}Y} \; \left[\int_{Y_0}^Y u_2(t) \; {\rm e}^{\int_{Y_0}^t u_1(t) \; {\rm d}t} {\rm d}t +R_2(0) \right].
\label{r2sa}
\end{eqnarray}

As in the case of $R_1$, the large $N$ limit of the above solution is again expected to be consistent with large $N$ solution obtained by rescaling of $Y$; this is  discussed in {\it appendix H}.

\subsection{Connection of complexity parameter with time}

As mentioned in section I, a variation of system parameters  in time subjects the quantum state and thereby its  entanglement measures to evolve in time $t$. As a typical many body state is represented by an ensemble, and the  parameters  of an appropriate  ensemble must be functions of system parameters, this in turn leads to time evolution of the ensemble parameters; we note the latter may also contain an explicit time dependence. Consider a bipartite system with $g$ system parameters, say, $x_1(t), \ldots, x_g(t)$, with its quantum state represented by the ensemble (\ref{rhoc}) with 
$$h_{kl} = h_{kl}(x_1, \ldots, x_g, t), \quad b_{kl}=b_{kl}(x_1, \ldots, x_g, t).$$
 We then have $\frac{d h_{kl}}{dt}=\sum_{n=1}^g \frac{\partial h_{kl}}{\partial x_n} \; \frac{d x_n}{dt} + \frac{\partial h_{kl}}{\partial t} $. Following from $Y=Y(\{h_{kl} \}, \{b_{kl}\})$, we then have
$$\frac{d Y}{dt}=\sum_{n=1}^g \sum_{k\le l} \frac{\partial Y}{\partial h_{kl}}  \frac{\partial h_{kl}}{\partial x_n} \; \frac{d x_n}{dt} + \frac{\partial Y}{\partial t} .$$ The above can be rewritten as $$\frac{d Y}{dt}=\sum_{n=1}^g \frac{\partial Y}{\partial x_n}   \; \frac{d x_n}{dt} +\frac{\partial Y}{\partial t}$$ and this leads  to a description of $Y$ as a functional of time: 
$Y=Y(x_1(t), \ldots, x_g(t), t)$. Clearly for the states where explicit time-dependence  of $x_1, \ldots, x_g$ is known, their substitution in the above relation  gives $Y$ as a function of time and thereby evolution of $\langle R_1 \rangle, \langle R_2 \rangle$ with time. As a simple example, we consider the case with only one system parameter, say $x$ varying with time i.e $g=1$ and the state described by the ensemble parameters as 

\begin{eqnarray}
 b_{kl}=0, \; h_{kk}=1, \; h_{kl}=f(x(t)) \;  {\rm for} \; k\not=l
\label{sim}
\end{eqnarray}

The above relations along with eq.(\ref{y}) then gives
$\frac{d Y}{dt} \approx \frac{1}{1-2\gamma f}  \frac{\partial f}{\partial x} \; \frac{d x}{dt}$.  
For a case with $f(x)=x$ and $x = t^{\alpha}$, we then have  $Y(t)= \alpha \,\int \frac{t^{\alpha-1}}{1-2\gamma t^{\alpha}} \; {\rm d}t ={\alpha\over 2\gamma} \log (1-2\gamma t^{\alpha}) \approx \alpha  \; t^{\alpha}$; (this can also be seen by a direct substitution of $h_{kl}=t^{\alpha}, h_{kk}=1, b_{kl}=0$  in eq.(\ref{y})).

Previous studies, based on systems with and without conservation laws,  indicate  that $R_1$ evolves ballistically with time $t$; this is attributed to the spreading of basis operators in a ball of radius $r \sim v \, t$, where $v$ is the Lieb-Robinson velocity. Contrary to $R_1$, time-growth of $R_2$ however depends on whether the system is subjected to additional constraints  besides Hermiticity. For example, for systems in which the conserved quantities like energy evolve diffusively, $\propto \sqrt{t}$, $R_2$ is believed to scale diffusively with time, $R_2 \sim \sqrt{t}$ \; \cite{rpv,zl}; the latter behavior has been attributed to a slow decay of the largest Schmidt eigenvalue that contributes significantly to $R_2$ \cite{rpv}. (Alternatively, it can also be explained by considering the decay of overlap of unitary evolution of the full system with that of the independent evolution of its two parts \cite{zl}).

As discussed in previous section and explained by a simple example above, our theoretical analysis, based on $Y$ formulation, indicates that  the growth rates of $R_1$ and $R_2$ not only depends on how the system conditions change with time but also on the structure of the ensemble representing them as well as on the functional dependence of the ensemble parameters on system parameters. For example, for some complex systems even without any conservation laws but described by the ensemble (\ref{sim}) with $Y \sim t^{\alpha}$, both $R_1, R_2$ can grow diffusively for small $t$ if $\alpha=1/2$ (or more generally as a power law).  Furthermore for the  quantum  states with reduced density matrix represented by a sparse $C$ ematrix nsembles, the ensemble parameters can have different time dependence, leaving $Y$ and thereby $R_1, R_2$ as a general function of time. In presence of the conservation laws, the  analysis is further complicated. This is because an appropriate ensemble of the state matrices  is then, in general, a structured Wishart ensemble (i.e an ensemble with matrix elements subjected to local constraints determined by conservation laws). As the local constraints in general lead to matrix element correlations, the $Y$ formulation  given by eq.(\ref{y}) is then no longer applicable.

\section{Numerical Verification of complexity parameter based  formulation of the entropies}

Based on the complexity parametric formulation,  different reduced matrix ensembles subjected to same global constraints e.g symmetry and conservation laws are predicted to undergo similar statistical  evolution of the Schmidt eigenvalues. The fundamental significance of the prediction makes it necessary to verify it by a detailed numerical analysis. In this section, we verify it by numerically  comparing the entanglement measures of  three different many body systems, represented by three multi-parametric Gaussian ensembles of real $C$ matrices with different variance types (with $ h_{kl} = \langle C_{kl}^2\rangle - (\langle C_{kl}\rangle)^2$ , $b_{kl} = \langle C_{kl}\rangle$). (We recall here that $C_{kl}$ corresponds to a component of the state $\Psi$ in product basis $|k l \rangle$ consisting of  eigenstate $|k \rangle$ of subsystem A and  $|l \rangle$ of subsystem B,  a choice of variance of $C_{kl}$ changing with $l$ implies change of correlation between two subunits). Further details are as follows (with schematic images given in figure 1)

\vspace{0.1in}

(i) {\bf Components with same variance along higher columns (EB):}   The ensemble parameters in this case are  same for all elements except those in first column: 

\begin{eqnarray}
h_{k1} =1,  \hspace{0.1in} h_{kl}  = \frac{ 1}{ (1+\mu)}, l\neq 1,  b_{kl;s} = 0 (\forall \; k,l).
\label{vrp}
\end{eqnarray}
 The substitution of the above in eq.(\ref{y}) leads to
 
\begin{eqnarray}
Y=-\frac{N(N-1)}{2M\gamma}\left[ {\rm ln} \left(1-\frac{2\gamma}{(1+\mu)} \right)\right] + c_0
\label{yrp}
\end{eqnarray}
with constant $c_0$ determined by the initial state of the ensemble. 

Choosing initial condition with $\mu \rightarrow \infty$ corresponds to  an ensemble of $C$-matrices with only first column elements as non-zero; this in turn gives $Y_0=c_0$.

\vspace{0.1in}

(ii) {\bf Components variance decaying as a Power law along columns (EP):} The variance of the $C_{kl}$ now changes, as a power law, across the column as well as row but its mean is kept zero:  

\begin{eqnarray}
h_{kl} = \frac{1}{1+\frac{k}{b}  \frac{(l-1)}{a}},\qquad  b_{kl}  =  0 \quad \forall \;\; k, l
\label{vpe}
\end{eqnarray}
where $a$ and $b$ are arbitrary parameters. Eq.(\ref{y}) then gives
 
\begin{eqnarray}
Y=-\frac{1}{2M\gamma}\left[ \sum_{r_1=1}^{N-1} \sum_{r_2=1}^{N}  {\rm ln} \left(1-\frac{2\gamma}{1+r_1 r_2/a b}\right)\right] + c_0
\label{ype}
\end{eqnarray}
Choosing initial condition with $b, a \rightarrow \infty$ again corresponds to  an ensemble of $C$-matrices with only first column elements as non-zero and thereby $Y_0=c_0$.

\vspace{0.1in}

(iii) {\bf Components variance with exponential decay along columns (EE): } Here again the mean is kept zero for all elements but the variance changes exponentially across the column as well as row:  

\begin{eqnarray}
h_{kl} = {\rm exp}\left(-\frac{k |l-1|}{a b}\right),\qquad  b_{kl} = 0 \qquad  \forall \;\; k, l
\label{vee}
\end{eqnarray}
with $b$ as an arbitrary parameter. Eq.(\ref{y}) now gives  

\begin{eqnarray}
Y=-\frac{1 }{2M\gamma}\left[ \sum_{r_1=1}^{N-1} \sum_{r_2=1}^{N} \; {\rm ln} \left(1-\frac{2\gamma}{{\rm exp}(\frac{r_1}{a} \frac{r_2}{b})}\right)\right]  + c_0
\label{ch28}
\end{eqnarray}
with $M=N^2$.
Here again the initial choice of parameters $b, a \rightarrow \infty$ leads  to  a $C$-matrix ensemble  same as in above two cases and same $Y_0$.

For numerical analysis of various entropies, we exactly diagonalize  (using Lapack subroutine for real matrices  based on Lanczos algorithm) each ensembles for many matrix sizes  and for many values of  the ensemble  parameters $a, b$ but with fixed $\gamma=1/4$.  We note that in each of the three ensembles (schematic illustration in figure 1 for visual guidance), the change of variance along the columns ensures the variation of entanglement from an initial  separable state to maximally entangled state. Here the separable state corresponds to very small $a, b$ (equivalently $Y \to Y_0$) and the maximally entangled state corresponds to the ensemble with large $a, b$ ($Y > 1$). For simplification, the $C$ matrix chosen for all cases is a $N \times N$ square matrix, thus implying $\nu_0=0$. 
The  obtained Schmidt eigenvalues are then used to numerically derive the von Neumann and R\'enyi entropies (using $\log \; base \; 2$) for each matrix of the ensemble under consideration.

Figures \ref{avgEntDynVN} and \ref{compBetasR2} illustrates the $Y$-governed evolution of the ensemble averaged  von Neumann as well as second order R\'enyi entropy for three ensembles and for two symmetry conditions, namely, the quantum state with/ without time-reversal symmetry, respectively.  As clear from these figures, although both entropies grow with increasing $Y$ (as expected due to $Y=Y_0$ corresponding to the separable limit and a large $Y$ to the maximum entanglement), their growth rate  follows a quantitatively different, although  qualitatively similar, routes with same functional dependence; this is consistent with our theoretical prediction too. (It must be emphasized here that same $Y_0$ for the above three ensembles does not by itself imply same values for other parameters i.e $y_2, \ldots, y_M$ in the complexity parameter set and therefore they need not fall on the same evolutionary path). More specifically, (i) both entropies show sensitivity to symmetry conditions for small $N$ (as can be seen from difference of $Y$ dependent growth for $\beta=2$ from $\beta=1$ depicted in figures \ref{compBetasR2}),  (ii) both entropies on average show almost linear increase, growing to a maximum limit dependent on the global constraints on the ensembles, (iii) the behavior in the intermediate region i.e $Y \sim 1$ however is dependent on the ensemble details that is local constraints: while the growth is rapid/ sudden in ensemble with constant coefficients for higher columns, it is slower and relatively smoother in the other two ensembles. We also find different growth rates of the two entropies, quantitatively, with $R_2$ always slower than $R_1$ for all three cases.

Our analysis is section III required information about the average behavior of two additional measures namely $\langle R_0 \rangle$ and $\langle {1\over S_2} \rangle$ (with $R_0=-\sum_n \log \lambda_n$ and $S_2 =\sum_n \lambda_n^2$) with respect to $Y$ as well as $N$. This motivates to numerically  analyze the above averages for the three ensembles; the results are displayed in figures \ref{avgLoglmda} and \ref{avgInvS2}, respectively. As the figure \ref{avgLoglmda} indicates, $\langle R_0 \rangle$ has a very weak $Y$-dependence and quickly approaches the stationary limit: $\langle R_0(Y) \rangle \propto N \log N$ for $Y > o(1/N)$. While the results in  figures $\ref{avgLoglmda}\,(a,b,c)$ correspond to  a single $N$ value i.e $N=1024$, our numerical analysis indicates their validity in general in large $N$ limit. Contrary to $\langle R_0 \rangle$, $\langle {1\over S_2} \rangle$ displayed in figure \ref{avgInvS2} varies rapidly with $Y$: $\langle {1\over S_2} \rangle \sim N \; (1- {\rm e}^{-s_0 Y})$ with $s_0$ dependent on the ensemble.  The parts $\ref{avgInvS2} \, (d)$ confirm the expected behavior in  large $Y$ limit (stationary limit): $\langle   {1\over S_2}  \rangle \propto N \log_2 N$; (here the results are displayed only for one of the ensembles, stationary limit of all ensembles being same).

The analysis is section III also required information about how the functions $R_1(\lambda)= -\sum_n \lambda_n \log \lambda_n$ and $R_2=-\log \sum_n \lambda_n^2$ vary with trace $S_1=\sum_n \lambda_n $ of the reduced matrix ensemble. Figure \ref{r2r1vsS1} depicts the above dependence  for a stationary Wishart ensemble; as clear from the figure, ${\partial R_1 \over \partial S_1} >0$ and ${\partial R_2 \over \partial S_1} < 0$.

\section{Discussion: physical interpretation of Complexity parameter}

As the complexity parameter $Y$ is an important ingradient of our  theoretical formulation, it is necessary to discuss its implications in context  of the reduced density matrix and the entanglement measures. The details of derivation of $Y$ from the ensemble parameter sets $\{h, b\}$ is discussed in {\it appendix A}. To gain a physical insight in the origin of $Y$,  we first deconstruct eq.(\ref{trho1}) and seek the physical interpretation  of each of its parts.

While a point in $h, b$-space corresponds to an ensemble of $C$-matrices, the one in the matrix space corresponds to a single $C$-matrix. With $\rho_c$ a function of ensemble parameters (i.e $h_{kl}, b_{kl}$) as well as matrix elements ($C_{kl}$), the generators $T$ and $L$ in eq.(\ref{trho1}) describe the dynamics in different spaces:  $T$ generates an evolution of $\rho_c$ in the $h,b$ space while $C_{kl}$ ($\forall k,l$) are kept fixed, $L$ describes it in the matrix space  (i.e $C_{kl}$ now varying) with a fixed set $h, b$. Thus eq.(\ref{trho1}) describes the dynamics when a small perturbation of $\rho_c$ in the matrix space can be exactly mimicked by the one in the ensemble parameter space. This is expected on grounds that the dynamics in both spaces originates by variation of the system parameters. Although the latter are not identified here (as we begin with an ensemble directly),  but as mentioned in section III.c, for an ensemble to appropriately represent a complex system statistically, its parameters must be derived from the system parameters, e.g., strength of local interactions, disorder, dimensionality, boundary conditions etc. As discussed in section III.c, a knowledge of relation between the two then gives $Y$ as a function of various system parameters.  

The $h,b \to y$ transformation, defined by eq.(\ref{hby}), implies that the evolution of $\rho_c$ in the matrix space can also be mimicked by the dynamics in the complexity parameter space (consisting of $y_1, \ldots, y_M$). As eq.(\ref{ch8}) indicates, the variation of $\rho_c$, depends on $h_{kl}$, $b_{kl}$ (for all $k,l$) only through a function $Y \equiv y_1(h,b)$ while $y_2(h,b), \ldots, y_M(h,b)$ remain fixed at their initial values. This encourages following implication:  the $y$-space can essentially be viewed as a  ''centre of mass'' frame of reference, with $y_1$ acting as a ''centre of mass'' of the ensemble parameters (corresponding to average uncertainty associated in determination of the matrix $C$). The remaining parameters $y_2,\ldots, y_M$, acting  as the relative distances, are basically the functions of uncertainties which remain constant during the evolution of the quantum state. 

An important implication of the above is following:  each point in the path defined by $y_1(h,b),\ldots, y_M(h,b)$ in $y$-space can represent many different $C$-ensembles and thereby different quantum states. Two states are predicted to share the same distribution of the Schmidt eigenvalues and thereby the entanglement entropy if they belong to same $y_k \forall k$, equivalently, if they have same $y_1$ and evolve from a statistically same initial state. The latter ensures same values of $y_2, \ldots, y_M$ and intuitively appears to be the case for the quantum states belonging to same class of global constraints e.g. symmetry, conservation laws; a rigorous analysis is however needed  in this context. The set of complexity parameters can then be used as a classification criteria of the  entanglement for quantum states.

The above insight gives rise to an important query: is it always possible to define a transformation of the set of $M$ non-zero  parameters $h$ and $b$   to another set $\{y_1,..,y_M\}$ i.e  $h_{kl}=h_{kl}(y_1,..,y_M)$ and $b_{kl}=b_{kl}(y_1,..,y_M)$ such that  $y_2,..,y_M$ remain constant?The answer lies in the type of the perturbation of the quantum state: if the perturbation, due to variation of system parameters, leaves at least  $M-1$ state matrix  constraints invariant, the latter, or their combinations, can then be identified with $y_2,\ldots, y_M$. For example, if the perturbation is analyzed in a fixed basis e.g. product basis,  $y_2, \ldots, y_M$ can be chosen as the functions of basis parameters (as  the number $M$ of ensemble parameters depends on the number of basis states). 


An important consequence of the complex parametric formulation is that it identifies the hidden, relevant driver  as well as constants of the evolution of the entanglement measures: variation of the system conditions can cause a change of entanglement measures of a typical state  only if $Y$ changes. It is interesting to note that  $Y$ has a form of an information theoretic function: it is a sum over logarithmic values of the uncertainties associated with pairwise correlations of  the basis-states and is therefore a measure of total information content, (also called surprise or self-information) of basis state correlations. It is therefore not surprising that $Y$ governs the evolution of entanglement entropies. Further, for the states subjected to same global constraints e.g symmetry and conservation laws,
the complexity parameter formulation predicts an analogous average evolution  of the entanglement measures. Intuitively this suggests the following: the underlying complexity of the system wipes out details of the correlations between two sub-bases, leaving their entanglement to be sensitive only to an average measure of the complexity  i.e $Y-Y_0$.  Besides fundamental significance, the formulation is therefore useful for the following reason too: for states evolving along the same path, $Y$ can  be used as a hierarchical criteria  even if they belong to different complex systems but same global constraints.

\vspace{0.1in}

\section{Conclusion}
In the end, we summarize with a brief discussion of our main idea, results and open questions. 
We have theoretically analyzed the  system-dependence of the evolution of the  average entanglement entropies of a typical quantum state in a bipartite basis. While previous theoretical studies have mostly considered ergodic states leading to a standard Wishart ensemble representation of the reduced density matrix, our primary focus has been on the non-ergodic states, specifically  the states with their component (in bipartite basis) as Gaussian distributed with arbitrary variances and mean values and subjected to symmetry constraints. This in turn gives the reduced density matrix as a multi-parametric Wishart ensemble with fixed trace and  permits an analysis of the entanglement  with changing system conditions, i.e, the complexity of the system. Supported by detailed numerical analysis presented here, our theoretical results for ensemble averaged von Neumann as well as second order R\'enyi entropies   indicate a linear behavior only for small $Y$-ranges i.e near separability limit. In large $Y$-limit, both entropies approach a constant value dependent on the global constraints, but the change from linear to constant behavior is exponentially rapid. The growth rates of the two entropies with changing complexity are however different. We also find the results to be sensitive to global symmetry conditions with different growth rates  of entanglement for the states with and without time-reversal symmetry as well sensitivity to subsystem ``cut''.

An important ingredient of our approach is the common mathematical formulation  of the  joint probability distribution of the Schmidt eigenvalues for different quantum states  which in turn leads to the evolution equations for the distributions of the  R\'enyi and von Neumann entropies in terms of a single function of all system parameters.  The formulation, referred as complexity parameter formulation,  is a generalization of a similar formulation reported in \cite{pslg, ptche} (for unconstrained Wishart and chiral ensembles) and not only leads to a technical simplification but also provides fundamental insights about the entanglement relations between different quantum states (including states belonging to different local system constraints). For states, where explicit time-dependence  of the system parameters  is known, the complexity parameter can be expressed as  a function of time and therefore can be used to derive information about the time-dependent evolution  of the entanglement measures.

The present study still leaves many open questions e.g. whether a knowledge of the average entanglement measures is sufficient for a typical state of a complex system or their fluctuations should also be taken into account? The consideration would have significant impact on any hierarchical arrangement of states. Another important question is regarding the explicit role played by the system parameters in the entanglement evolution. The complexity parameter depends on the system parameters through distribution parameters. A knowledge of their exact relation however  requires a prior knowledge of the quantum Hamiltonian, its matrix representation in the product basis as well as the nature and distribution parameters of the appropriate ensemble.  The knowledge leads to the determination of the appropriate    ensemble of its eigenstates. As indicated by recent studies, the conservation laws of a system can also affect the entanglement measures and their growth. This is turn requires a  detailed investigation of the entropies of reduced density matrices represented by structured/ correlated  Wishart ensembles. 

Another important query arises regarding the role of $T$, i.e, the particular combination of first order parametric derivatives, in $Y$ governed evolution of $\rho_c$ and thereby the entanglement measures.  The choice of $T$ permits a linear transformation from $h,b$-space to $y$-space resulting $Y$ in the form of a information theoretic function. This also ensures that the Gaussian densities spread linearly by the same rate and their mean values too shift in a same way, finally reaching to a stationary state corresponding to maximum entanglement.  It is then natural to wonder how far a Gaussian choice of $\rho_c$ is necessary for the complexity parameter formulation? For example, is it possible to  reach to same deductions by defining another $T$-operator if $\rho_c$ is non-Gaussian?  We intend to answer some of these queries in near future.

\acknowledgments

One of the authors (P.S.)  is grateful to SERB, DST, India for the financial support provided for the  research under Matrics grant scheme.

\appendix

\section{Complexity parameter formulation}

A perturbation of the state by a change of the parameters $h_{kl;s} \rightarrow h_{kl;s}+\delta h_{kl;s}$ 
and $b_{kl} \rightarrow b_{kl}+\delta b_{kl}$  over time causes the matrix elements $C_{kl;}$ undergo a dynamics in the matrix space. We consider a combination of multiparametric variations defined as  

\begin{eqnarray}
T \; \rho_c \equiv \sum_{k\le l;s}\left[(2/\tilde g_{kl}) x_{kl;s}{\partial \rho_c\over\partial h_{kl;s}} - \gamma
  b_{kl;s} {\partial \rho_c \over\partial b_{kl;s}}\right]
 \label{trho}
\end{eqnarray}   
For such a combination along with its Gaussian form,  the multi-parametric evolution of $\rho_c(C)$ can then be described exactly  in terms of a diffusion, with finite drift,  in $C$-matrix space: 

\begin{eqnarray}
T \rho_c =  L \rho_c + (T \log {\mathcal N}) \; \rho_c
\label{trho1}
\end{eqnarray}
with

\begin{eqnarray}
L \rho_c =  \sum_{k,l,s} \frac{\partial}{\partial C_{kl;s}}\left[ \frac{\partial  \rho_c}{\partial C_{kl;s}}+\gamma \; C_{kl;s} \; \rho_c \right].
\label{lrho1}
\end{eqnarray}

As the above equation  is difficult to solve for generic parametric values,  we seek a transformation from the set of $M$ parameters $\{h_{kl;s}, b_{kl;s} \}$ to another set $\{y_1,\ldots, y_M \}$
such that only $y_1$ varies under the evolution governed by the operator $T$ and rest of them i.e $y_2, \ldots, y_M$ remain constant: 
$T \rho_c \equiv \frac{\partial \rho_c}{\partial y_1}$.  This in turn requires 

\begin{eqnarray}
T y_1=1, \qquad T y_k=0  \qquad \forall \; \;  k >1. 
\label{hby}
\end{eqnarray}
Using $T$ given by eq.(\ref{trho}), the above set of equations can be solved by the standard method of characteristics

\begin{eqnarray}
  \frac{d h_{kk;s}}{ x_{kk;s}} &=& \ldots = \frac{d h_{kl;s}}{2 x_{kl;s}} = \frac{db_{kl;s}}{b_{kl;s}} = {dy_1} \label{lt1} \\
  \frac{d h_{kk;s}}{ x_{kk;s}} &=& \ldots = \frac{d h_{kl;s}}{2 x_{kl;s}} = \frac{db_{kl;s}}{b_{kl;s}} =\frac {dy_n}{0}  \qquad  n>1
\label{ltn}
\end{eqnarray}
where equality relations include all $\{kl;s \}$ pairs. The solution of eq.(\ref{lt1}) is  $y_1 = Y$ with $Y$ given by eq.(\ref{y}) and rest of the equations give $y_k=c_k$ with $c_2, \ldots,c_M$ as the constants of evolution. 


By defining $\rho = {\rm e}^{\int f_0 {\rm d}y_1} \; \rho_c$ with $f_0=  T \log {\mathcal N} = \sum_{kl;s} {1- \gamma h_{kl;s} \over 2 h_{kl;s}}$,  eq.(\ref{lrho1}) can be written as the  diffusion equation for $\rho$ 

\begin{eqnarray}
\frac{\partial \rho}{\partial y_1} =  \sum_{k,l,s} \frac{\partial}{\partial C_{kl;s}}\left[ \frac{\partial  \rho}{\partial C_{kl;s}}+\gamma \; C_{kl;s} \; \rho \right]
\label{ch8a}
\end{eqnarray}

\section{Diffusion of reduced density matrix $\rho_A$}
	    
As expected, the diffusive dynamics of the matrix elements $C_{kl}$ manifests itself in the $\rho_A$-matrix space and the moments for  the matrix elements $\rho_{A; mn}= \sum_{k=1}^{N_a} C_{mk} C_{nk}^*$ can be calculated from those  of $C$. As discussed in \cite{sp,pslg},  eq.(\ref{att0}) can be rewritten as

\begin{eqnarray}
C(Y+\delta Y)  &\equiv&  {C(Y) + \sqrt{2 \; \delta Y} \; V(Y) \over \sqrt{1 + 2 \; \gamma \; \delta Y} } \label{att} \\
&\approx &  C(Y) \; \left(1 -  \gamma \; \delta Y \right) +  \sqrt{2 \; \delta Y} \; V(Y) + O((\delta Y)^{3/2}). 
\label{atm1}
\end{eqnarray} 
Here the symbol $''\equiv''$ implies the equivalence of the ensembles of matrices on two sides.   
The ensemble  approaches to equilibrium as $Y \to \infty$. The equivalence of  eq.(\ref{att0}) and eq.(\ref{att}) along with the derivation of the diffusion equation for $C(Y)$ is 
discussed in \cite{sp}. 

As expected, the diffusive dynamics of the matrix elements $C_{kl}$ manifests itself in the $\rho_{A}$-matrix space and the moments for  the matrix elements $\rho_{A; mn}= \sum_{k=1}^{N_a} C_{km}^* C_{kn}$ can be calculated from those  of $C$. The above equations along with  relation between the elements of $\rho_{A}$ and $C$ gives the  moments of the matrix elements of $L$. As discussed in appendix A of \cite{pslg}, the expression for $1^{st}$ moment is of the same form for both $\beta=1$ or $2$: 

\begin{eqnarray}
\langle{\delta \rho_{A; mn}} \rangle &=&  2 \; (\beta \; v^2 \; N_a \; \delta_{mn} -  \gamma \; \rho_{A; mn}  )\; \delta Y 
\label{lmn0}
\end{eqnarray}
but the terms in  $2^{nd}$ moment differ  as follows

\noindent{\bf Case $\beta=1$} 

\begin{eqnarray}
\langle\delta \rho_{A;mn} \; \delta \rho_{A; kl}^* \rangle  &=& \langle\delta \rho_{A; mn} \; \delta \rho_{A; kl} \rangle \nonumber \\
&=& 2 \; v^2 \; [\rho_{A; mk}  \; \delta_{nl} +\rho_{A; ml}  \; \delta_{nk} + \rho_{A; nk} \; \delta_{ml} + \rho_{A; nl} \; \delta_{mk}] \; \delta Y
\label{lmn1}
\end{eqnarray}

\noindent{\bf Case $\beta=2$} 

\begin{eqnarray}
\langle\delta \rho_{A; mn} \; \delta \rho_{A; kl}^* \rangle  
&=& 4 \; v^2 \; [\rho_{A; mk} \; \delta_{nl}  + \rho_{A; nl}^* \; \delta_{mk}] \; \delta Y \nonumber \\
\langle\delta \rho_{A; mn} \; \delta \rho_{A; kl} \rangle  
&=& 4 \; v^2 \; [\rho_{A; ml}  \; \delta_{nk}  + \rho_{A; nk}^* \; \delta_{ml}] \; \delta Y
\label{lmn2}
\end{eqnarray}
 with $\beta=1, 2$  for $\rho_{A}$ real symmetric or complex Hermitian, respectively. 

\section{Diffusion of Schmidt eigenvalues}
	    
The diffusion of the matrix elements of $\rho_A$ manifests itself in the dynamics  
 of its eigenvalues and eigenfunctions. The  evolution equation for the  joint probability density  function (JPDF) of all eigenvalues  can now be derived as follows.

 Let $U$ be  the $N\times N$ eigenvector matrix of $\rho_A(Y)$ , unitary in nature i.e $U^{\dagger}  U=1$ ($L$ being  Hermitian) and $\lambda$ be the $N\times N$ diagonal matrix of  its eigenvalues,  $\lambda_{mn}= \lambda_n \; \delta_{mn}$. A small change $\delta Y$ in parameter $Y$ changes $\rho_A$ and its eigenvalues  and eigenfunctions. Using standard 
perturbation theory for Hermitian operators and by considering matrix $\rho_A+\delta \rho_A$ in the eigenfunction representation of matrix $W$,  
a small change $\delta \lambda_n$ in the eigenvalue $\lambda_n$ can be given as 

\begin{eqnarray}
\delta \lambda_n = \delta \rho_{A;nn} +\sum_{m\not=n} {|\delta \rho_{A; mn}|^2 \over \lambda_n-\lambda_m}+
o((\delta \rho_{A;mn})^3)
\label{den}
\end{eqnarray}
where $\rho_{A;mn}=\lambda_n \; \delta_{mn}$ at value $Y$ of complexity parameter (due to $\rho_A+\delta \rho_A$ being considered in the diagonal representation of $\rho_A$). Eq.(\ref{den}) gives, up to first order of $\delta Y$ (see {\it appendix} B),   

\begin{eqnarray}
\langle{\delta \lambda_n} \rangle
&=& 2\beta v^2 \; \left[  N_a -  {\gamma\over \beta v^2} \; \lambda_n +   \sum_{m=1,m\not=n}^{N} 
\; {\lambda_n + \lambda_m \over \lambda_n-\lambda_m}\right] \delta Y 
\label{enm0} \\
\langle{\delta \lambda_n \delta \lambda_m }\rangle &=& 
 8\; v^2 \; \lambda_n \; \delta_{nm} \; \delta Y 
\label{enm1}
\end{eqnarray}

In general, assuming Markovian process, the parametric diffusion of the  joint probability distribution $P_x(x_1,\ldots, x_N; Y)$ of $N$ variables $x_n$, $n =1,\ldots, N$ from an arbitrary initial condition, with $Y$ as the parameter, is given by the standard Fokker-Planck equation

\begin{eqnarray}
{\partial P_x\over\partial Y} \; \delta Y = {1\over 2} \sum_{k,l=1}^N {\partial^2 \over \partial x_{k} \partial x_l} \; (\langle\delta x_k \delta x_l \rangle\; P_x)  -\sum_{k=1}^N  {\partial \over \partial x_k} \; (\langle \delta x_k \rangle \; P_x)
\label{px}
\end{eqnarray}
Using  the above, the diffusion equation for the joint probability density 
for $P_{ev} ( \{\lambda_n \}; Y)$ at perturbation strength $Y$ where $\{\lambda_n\}$ refer to the sets of all  eigenvalues $\lambda_1, \lambda_2,..,\lambda_N$ can now be given as

\begin{eqnarray}
 {\partial P_{\lambda} \over\partial Y}\; \delta Y &=& \sum_{n} {\partial \over \partial \lambda_n}\left[{1\over 2} \; {\partial \over \partial \lambda_n} \langle (\delta \lambda_n)^2 \rangle- \langle \delta \lambda_n \rangle \right]  P_{\lambda}
 \label{pdl0}
 \end{eqnarray}

Note here $P_{\lambda}$ is subjected to following boundary condition: $P_{\lambda} \to 0$ for $\lambda_n \to [0, \infty )$ for $n=1 \to N$; this follows because the higher order moments of the ensemble density are assumed to be negligible. Substitution of the moments from eq.(\ref{enm0}) and eq.(\ref{enm1}) lead to the  diffusion equation for   $P({\lambda})$  given by eq.(\ref{pdl1}).

\section{Derivation of eq.(\ref{i1})}

For$f(\lambda)$ as an arbitrary function of eigenvalues $\lambda_1 \ldots \lambda_N$, $I_1$ in eq.(\ref{i1}) can be written as

\begin{eqnarray}
I_1 &=& - C_{hs} \; \sum_{n=1}^N \int_0^{\infty}   \;  \delta_1 \; f(\lambda) \;  \frac{\partial}{\partial \lambda_n}\left( \sum_{m=1}^N \frac{\beta \lambda_n}{\lambda_n- \lambda_m} - {\beta \nu} - {2 \gamma} \lambda_n \right) \;  P_{\lambda} \; {\rm D} \lambda.
\label{i2a}
\end{eqnarray}
with $ P_{\lambda}(\lambda) \to 0$ at the two integration limits  $0 \le \lambda_n \le \infty$ $\forall n=1 \to N$. We note here that the constraint $\delta_1$ effectively reduces the limits to ${1\over N} \le \lambda_n \le \infty$. 

Integration by parts now gives $I_2= A +B$ where

\begin{eqnarray}
A &=&  C_{hs} \; \sum_{n=1}^N \int_0^{\infty}    \;  \left[\frac{\partial \delta_1}{\partial \lambda_n}  \; f   \right] \;  \left( \sum_{m=1}^N \frac{\beta \lambda_n}{\lambda_n- \lambda_m} - {\beta \nu} - { 2\gamma} \lambda_n \right)  P_{\lambda}  \; {\rm D} \lambda  \\
B &=& C_{hs} \; \sum_{n=1}^N \int_0^{\infty}      \left[  \delta_1 \; \frac{\partial f }{\partial \lambda_n}  \right] \;  \left( \sum_{m=1}^N \frac{\beta \lambda_n}{\lambda_n- \lambda_m} - {\beta \nu} - {2 \gamma} \lambda_n \right)  P_{\lambda} \; {\rm D} \lambda
\end{eqnarray}
Again using $\frac{\partial \delta_1}{\partial \lambda_n}  = - \frac{\partial \delta_1}{\partial S_1}$, $A$ can be rewritten as 

\begin{eqnarray}
A &=& -    C_{hs} \;\frac{\partial }{\partial S_1}  \int  \;  f \;  \delta_1 \;  \sum_{n=1}^N\left( \sum_{m=1}^N \frac{\beta \lambda_n}{\lambda_n- \lambda_m} - {\beta \nu} - { 2 \gamma} \lambda_n \right)  P_{\lambda} \; {\rm D} \lambda \\
&=&   \frac{\partial   }{\partial S_1} \left(2  \; \gamma \; S_1 - {1\over 2} \beta N(N-2\nu-1) \right) \; \langle f \rangle  
\label{i2aa}
\end{eqnarray}
with 
\begin{eqnarray}
\langle f\rangle \equiv \langle f(S_1) \rangle &=& C_{hs} \; \int    \delta_1 \; f \; P_{\lambda} \; {\rm D} \lambda \\
J \equiv J(S_1) &=&  C_{hs} \; \int  {\rm D} \lambda \;  \delta_1 \; P_{\lambda} 
\end{eqnarray}

Similarly 
\begin{eqnarray}
B =   \beta \; J_2 - C_{hs} \; \int    \delta_1 \; \left[ \sum_{n=1}^N \frac{\partial f }{\partial \lambda_n}  \; \left({\beta \nu} + { 2\gamma} \lambda_n \right)  P_{\lambda}\right] \; {\rm D} \lambda 
\label{i2ab}
\end{eqnarray}
where

\begin{eqnarray}
J_2 \equiv J_2(S_1) &=& C_{hs} \; \int  \;   \delta_1 \; \left[\sum_{m, n=1}^N \frac{\partial f }{\partial \lambda_n}  \; \frac{ \lambda_n}{\lambda_n- \lambda_m}\right] \; P_{\lambda} \; {\rm D} \lambda 
\label{i3ad}
\end{eqnarray}

Based on the details of function $f(\lambda)$, the integrals $A$ and $B$ can further be reduced. Here we give the results for $R_1$ and $R_2$.

{\bf Case $f=R_1(\lambda)$:} Following from above, we have for $f =-\sum_n \lambda_n \log \lambda_n$, 

\begin{eqnarray}
A &=&   \frac{\partial   }{\partial S_1} (2  \; \gamma \; S_1-{1\over 2} \beta N(N-2\nu-1)) \; \langle R_1 \rangle  \label{i2aaa}
\end{eqnarray}
and

\begin{eqnarray}
B  &=& -  C_{hs} \; \int  \; \delta_1 \; \left[ \sum_{m,n=1}^N \frac{\beta \lambda_n \left(1+\log \lambda_n \right)}{\lambda_n- \lambda_m} - {\beta \nu} \sum_n \left(1+\log \lambda_n \right) - { 2\gamma } \sum_n \lambda_n \left(1+\log \lambda_n \right) \right] \; P_{\lambda}  {\rm D} \lambda  \nonumber \\
\label{i2aba}
\end{eqnarray}
Eq.(\ref{i2aba}) can be rewritten as 

\begin{eqnarray}
B  &=& - \beta \; J_{3} - {1\over 2} \beta N(N- 2\nu -1) J -  {\beta \;\nu} \; \langle R_0 \rangle  + {2 \gamma} \; \left(S_1 \; J + \langle R_1 \rangle \right) 
\end{eqnarray}
where $R_0(\lambda) \equiv -\sum_n \log \lambda_n $ and 
$J_{3} = -  C_{hs} \; \int  \; \delta_1 \; \left[ \sum_{m,n=1}^N \frac{\beta \lambda_n \left(1+\log \lambda_n \right)}{\lambda_n- \lambda_m}  \right] \; P_{\lambda}  {\rm D} \lambda$.

To reduce $B$ further, we note that $|\lambda_n -\lambda_m| \le 1$; this in turn permits the approximation  
$\log \lambda_n \approx \log \lambda_m$ and thereby leads to the  relation 
 (derived in {\it appendix G})
\begin{equation}
\sum _{n \neq m} \frac{\lambda_n \log \lambda_n}{\lambda_n - \lambda_m} 
\approx \frac{N (N-1)}{2} - \frac{(N-1)}{2} R_0
\label{llr}
\end{equation}

Substitution of the above relation gives $J_3 \approx  \frac{N (N-1)}{2} \; J - \frac{(N-1)}{2} \langle R_0 \rangle$. This along with eq.(\ref{i2ac}) then leads to
\begin{eqnarray}
B  &=& -{1\over 2} \beta N(N- 2\nu -1) J - \frac{\beta}{2} N (N-1) + {1\over 2} \beta \; (N- 2\nu-1) \; \langle R_0 \rangle \nonumber \\ &+& {2 \gamma} \; \left(S_1 \; J - \langle R_1 \rangle \right) 
\label{i2ac}
\end{eqnarray}
Substitution of eq.(\ref{i2aaa}) and eq.(\ref{i2ac}) in eq.(\ref{i2a}),  we have for $f= R_1$,
\begin{eqnarray}
I_1 &=& -{1\over 2} \beta N(N-2\nu-1) J - \frac{\beta}{2} N (N-1) + {1\over 2} \beta \; (N-2\nu-1) \; \langle R_0 \rangle  + {2 \gamma} \; S_1 \; J \nonumber \\
&+&  (2  \; \gamma \; S_1-{1\over 2} \beta N(N-2\nu-1)) \; \frac{\partial  \langle R_1 \rangle }{\partial S_1}
\label{i2ci}
\end{eqnarray}

\vspace{0.5in}

{\bf Case $f=R_2$:} Proceeding similarly for $f= R_2 = - \log \sum_n \lambda_n^2$, we have, 
from eq.(\ref{i2aa}) and eq.(\ref{i2ab}),

\begin{eqnarray}
A &=&  \frac{\partial   }{\partial S_1} (2  \; \gamma \; S_1-{1\over 2} \beta N(N-2\nu-1)) \; \langle R_2 \rangle  
\end{eqnarray}
and

\begin{eqnarray}
B &=&   -2 C_{hs} \; \int    \delta_1 \;  {1\over S_2}  \left( \sum_{m, n=1}^N \frac{\beta \lambda_n^2}{\lambda_n- \lambda_m} - {\beta \nu} \sum_{n=1}^N \;\lambda_n - { 2\gamma} \sum_{n=1}^N \;\lambda_n^2  \right)  P_{\lambda} \; {\rm D} \lambda 
\end{eqnarray}
Rewriting $2 \; \sum_{m, n=1}^N \frac{\lambda_n^2}{\lambda_n- \lambda_m} = 
\sum_{m, n=1}^N \frac{\lambda_n^2-\lambda_m^2}{\lambda_n- \lambda_m} = 2 (N-1) \sum_n \lambda_n$, the above equation can be rewritten as 

\begin{eqnarray}
B &=&  -C_{hs} \; \int    \delta_1 \;   {1\over S_2}   \left( 2( N-1 -\nu) \beta \; S_1 - { 4\gamma} \; S_2 \right)  P_{\lambda} \; {\rm D} \lambda
\label{i2ae}
\end{eqnarray}

substitution of the above in eq.(\ref{i2a}) gives 

\begin{eqnarray}
I_1 &=& -2 \; \beta (N-\nu-1) S_1 \langle {1\over S_2} \rangle + 2 \gamma \;(\langle R_2 \rangle + 2 J) +   (2  \; \gamma \; S_1-{1\over 2} \beta N(N-2\nu-1)) \; \frac{\partial  \langle R_2 \rangle }{\partial S_1} \nonumber \\
\label{i2di}
\end{eqnarray}

\section{Derivation of eq.(\ref{i2})}

For$f(\lambda)$ as an arbitary function of eigenvalues $\lambda_1 \ldots \lambda_N$, $I_2$ defined in eq.(\ref{i2}) can be written as

\begin{eqnarray}
I_2 &=& C_{hs} \; \sum_{n=1}^N \int  \;  \delta_1 \; f(\lambda) \;  \frac{\partial^2 (\lambda_n \; P_{\lambda})}{\partial \lambda_n^2}  \;{\rm D} \lambda \\
&=&  C_{hs} \;\sum_{n=1}^N \int   \;  \left[\; \delta_1 \; \frac{\partial^2  f}{\partial \lambda_n^2}  \; \lambda_n  + 
2 \frac{\partial  \delta_1}{\partial \lambda_n} \; \frac{\partial f}{\partial \lambda_n} 
 \; \lambda_n + \frac{\partial^2  \delta_1}{\partial \lambda_n^2} \; f \; \lambda_n \; \right] \;   P_{\lambda} \; {\rm D} \lambda
 \label{i1a}
\end{eqnarray}

Now using $\frac{\partial \delta_1}{\partial \lambda_n}  = - \frac{\partial \delta_1}{\partial S_1}$ the above can be rewritten as

\begin{eqnarray}
I_2 &=& C_{hs} \; \int  {\rm D} \lambda \;  \delta_1 \; P_{\lambda} \; F_1 - 
 2  \frac{\partial }{\partial S_1}  \int  {\rm D} \lambda \;  \delta_1 \; P_{\lambda} \; F_2
 +  \frac{\partial^2 }{\partial S_1^2}  (S_1 \; \langle f \rangle) 
 \label{i1b}
\end{eqnarray}
where $F_1(f)=\sum_{n=1}^N \frac{\partial^2  f}{\partial \lambda_n^2} \; \lambda_n$ 
and $F_2(f)=\sum_{n=1}^N  \frac{\partial  f}{\partial \lambda_n}  \; \lambda_n $ and last term in eq.(\ref{i1b}) is obtained by the relation 

\begin{eqnarray}
C_{hs} \; \int  {\rm D} \lambda \;  \delta_1 \; P_{\lambda} \; \left[ \sum_{n=1}^N \lambda_n \right] \; f
= C_{hs} \; S_1 \; \int  {\rm D} \lambda \;  \delta_1 \; P_{\lambda} \;  f = S_1 \; \langle f \rangle.
\label{ii1b}
\end{eqnarray}

{\bf Case $f=R_1$:}  substituting $f=-\sum_n \lambda_n \log \lambda_n$ in eq.(\ref{i1b}) and using $F_1=-\sum_n {\lambda_n \over \lambda_n}=-N$ and $F_2=-\sum_{n=1}^N (\log \lambda_n +1) \lambda_n$, 
$I_2$ can be written as 

\begin{eqnarray}
I_2 &=& -N J + 2  \; \frac{\partial  }{\partial S_1}  \left( S_1 \; J - \langle R_1 \rangle \right) 
+  \frac{\partial^2 }{\partial S_1^2}  (S_1 \langle R_1 \rangle) \\
&=& -(N-2) J + S_1 \; \frac{\partial^2 \langle R_1 \rangle  }{\partial S_1^2}  
\label{i1c}
\end{eqnarray}
where $\langle R_1 \rangle \equiv \langle R_1(S_1) \rangle $ and $J \equiv J(S_1) = \int  {\rm D} \lambda \;  \delta_1 \; P_{\lambda}$. 

{\bf Case $f=R_2$:}   Proceeding similarly for $f= R_2 = - \log \sum_n \lambda_n^2$, we have (now $\langle R_2 \rangle \equiv \langle R_2(S_1) \rangle $)

\begin{eqnarray}
I_2 &=& 4 \langle {S_3 \over S_2^2} \rangle - 2 S_1 \langle {1\over S_2} \rangle 
+ 4 \frac{\partial  J}{\partial S_1} 
+ \frac{\partial^2  }{\partial S_1^2}  (S_1 \; \langle R_2 \rangle) \\
&=& 4 \langle {S_3 \over S_2^2} \rangle - 2 S_1 \langle {1\over S_2} \rangle
+ 4  \; \frac{\partial J }{\partial S_1} + 2\frac{\partial \langle R_2 \rangle }{\partial S_1} 
+ S_1 \; \frac{\partial^2  \langle R_2 \rangle}{\partial S_1^2}  
\label{i1d}
\end{eqnarray}

\section{Solutions of eq.(\ref{vnd1}) and eq.(\ref{rn2})}

We consider following differential equation
\begin{equation}
  {1\over a} \;  \frac{\partial F}{\partial \Lambda} + \frac{\partial F}{\partial S_1}= S_1^{\gamma} \; g(\Lambda, S_1),
  \label{df1}
\end{equation}
with $g$ as an arbitrary function of $\Lambda$ and $S_1$ and  $\gamma$ an arbitrary constant. Using the method of characteristics, the  solution of the above equationcan be obtained by the solving following relation
\begin{equation}
  \frac{d\Lambda}{1/a} = \frac{dS_1}{1} = \frac{dF}{ g \; S_1^{\gamma}}.
  \label{charcEqn}
\end{equation}
To solve the above, we rewrite the first equality as $\frac{d\Lambda}{dS_1} = \frac{1}{a} $ which on solving gives
\begin{eqnarray}
\Rightarrow \Lambda &=& \frac{1}{a}S_1 + C_0.
\label{LambdaS}
\end{eqnarray}
with $C_0$ as a constant of integration. Replacing $\Lambda$ by the relation in eq.(\ref{LambdaS}), the second equality in eqn. \eqref{charcEqn} gives $\frac{dF}{dS_1} = S_1^{\gamma} \, g(\Lambda(S_1), S_1)$.  Solving the latter then leads to $F - \int g \; S_1^{\gamma} \; {dS_1} =C_1$ with $C_1$ as a constant of integration. The general solution 
of eq.(\ref{df1}) or eq.({charcEqn}) can then be given as $C_1=C_2(C_0)$ or alternatively

\begin{eqnarray}
F = \int g \; S_1^{\gamma} \; {dS_1}  +  C_2\left(\Lambda - \frac{S_1}{a}\right),
\label{rra}
\end{eqnarray}
 where  $C_0$ is substituted from eqn. \eqref{LambdaS}. The latter can be determined by a  substitution of the above solution in eq.(\ref{df1}); it leads to
\begin{equation}
\frac{\partial C_2}{\partial (a \Lambda)} =- \frac{\partial C_2}{\partial S_1}.
\label{df2}
  \end{equation}
Solving the above by separation of variables method now gives $C_2\left(\Lambda - \frac{S_1}{a}\right) = B - e^{\tau\,(S_1-a\,\Lambda)}$ with $\tau$ as an arbitrary constant. The solution of eq.(\ref{df1}) for arbitrary boundary condition can now be given as 
\begin{equation}
F=  B + \int g \; S_1^{\gamma} \; {dS_1}  - e^{\tau \,(S_1-a\,\Lambda)}
\label{ff}
\end{equation}
with $B$ and $\tau$  as unknown constants, to be determined from the boundary conditions. 

We note that a substitution of $\gamma=0, g(\Lambda, S_1)= \left({\langle R_0 \rangle\over N}-q_0 \, J\right)$  in eq.(\ref{df1})  gives eq.(\ref{vnd1}); the same substitution in eq.(\ref{ff}) then gives its solution (\ref{r1s2}).
Similarly the substitution of $\gamma=1, g(\Lambda, S_1)=-\eta \; \langle \frac{1}{S_2}\rangle$ in eq.(\ref{df1}) and eq.(\ref{ff})  gives eq.(\ref{rn2}) and  its solution (\ref{rn2a}) respectively.

\section{Solution eq.(\ref{r1sa}) in large $N$ limit}

As $v_2(Y) \approx  {1\over 2} \beta N_{\nu} \; \langle R_0 \rangle $, $v_1(Y) \approx -\phi(1) \approx {1\over 2}\beta N N_{\nu} g'(1)$ in the limit, eq.(\ref{r1sa}) can  be rewritten as 

\begin{eqnarray}
\langle R_1 (Y)\rangle \approx  -{ \beta \; N_{\nu} \over 2 }  \; {\rm e}^{\phi(1) \; (Y-Y_0)  } \; \int_{Y_0}^Y \langle R_0 \rangle \; {\rm e}^{-\phi(1) (t-Y_0) } \; {\rm d}t +\langle R_1 (0)\rangle \; {\rm e}^{\phi(1) (Y-Y_0)} . 
\end{eqnarray}

Based on our numerical analysis of three different ensembles (details discussed in section IV), we find that, for $Y > o(1/N)$ range,  $\langle R_0 \rangle$ rapidly becomes almost constant with respect to $Y$  (numerical results displayed in figure 4). The above can then be approximated as 
\begin{eqnarray}
\langle R_1 (Y)\rangle \approx {\langle R_0 \rangle \over N \; g'(1)}  \; \left(1-{\rm e}^{-{1\over 2} \; \beta N N_{\nu} g'(1) (Y-Y_0)}\right) +\langle R_1 (Y_0)\rangle \; {\rm e}^{-{1\over 2} \; \beta N N_{\nu} g'(1) \; (Y-Y_0)}
\label{r1avna}
\end{eqnarray}
Here again for  separable state chosen as initial one, we have $\langle R_1 (0)\rangle =0$.  For small $Y$ (but  $> o(1/N)$), the above equation then given a linear increase of the entropy with $Y$:
\begin{eqnarray}
\langle R_1 (Y)\rangle \approx   {1\over 2} \; {\beta \; N_{\nu} \; \langle R_0 \rangle \; (Y-Y_0)}.
\label{rvnsy}
\end{eqnarray}
 For large $Y$, the growth however is again rapid, with $\langle R_1 (Y)\rangle$ approaching a constant value. The $Y \to \infty$ limit of the above solution gives 
\begin{eqnarray}
\langle R_1 (\infty)\rangle ={v_2 \over v_1}  \approx {  \langle R_0 \rangle \over N  g'(1)}
\label{rvinf}
\end{eqnarray}
Our numerical analysis based on three different ensembles indicates $\langle R_0 \rangle \sim N \log N$ for large $Y$ and large $N$ ((displayed in figure 4, details given in section IV). This in turn gives expected limit $\langle R_1 (\infty)\rangle \sim \log N$ for a maximally entangled state; (as $\lambda_n \approx {1\over N}$ in large $Y$ limit, the logarithmic $R_1$ behavior is consistent with its expectation).

\section{Solution eq.(\ref{r2sa}) in large $N$ limit}

As in the case of $\langle R_2 \rangle$, here again we consider a  large $N$-limit of the solution in eq.(\ref{r2sa}). As ${1\over N^2} \le \langle S_3 \rangle \le 1$, we have $u_2(Y) \approx  - 2 \beta N \langle {1\over S_2} \rangle$, $u_1(Y) \approx -\phi(1) \approx  2 \;\beta N \; N_{\nu} \; g'(1)$. The above solution can now be approximated as 

\begin{eqnarray}
\langle R_2 (Y)\rangle \approx  - 2 \beta \; N \; {\rm e}^{\phi(1) \; (Y-Y_0)  } \; \int \langle {1\over S_2} \rangle \; {\rm e}^{-\phi(1) (t-Y_0) } \; {\rm d}t +\langle R_2 (0)\rangle \; {\rm e}^{\phi(1) (Y-Y_0) }. \nonumber \\
\label{r2avna}
\end{eqnarray}
It is important to note here that the term $\langle {1\over S_2} \rangle =\langle {\rm e}^{R_2} \rangle$ varies rapidly with $Y$, for small $Y$, and can not be treated as constant while integrating over $Y$. Baesd on our numerical analysis of three different ensembles (discussed in section IV and displayed in figure $5$), we  find $\langle {1\over S_2} \rangle  \sim N \; (1- {\rm e}^{-s_0 Y})$ with parameter  $s_0$ dependent on the ensemble.  For initial state chosen as separable, we have $\langle R_2 (Y_0)\rangle =0$.  For small $Y-Y_0$, the above equation then gives  (approximating $\langle {1\over S_2} \rangle  \sim N \; s_0 \; (Y-Y_0) $)

\begin{eqnarray}
\langle R_2 (Y)\rangle \approx  {s_0 \; N \; (Y-Y_0) \over |g'(1)| \; N_{\nu}}
\label{r2sya}
\end{eqnarray} 

For large $Y-Y_0$, however, the growth rapidly approaches a constant value. The $Y \to \infty$ limit of eq.(\ref{r2sa}) gives 

\begin{eqnarray}
\langle R_2 (\infty)\rangle ={u_2 \over u_1} = {\left(2\beta (N-\nu-1)+2\right) \; \langle {1\over S_2} \rangle - 4 \langle {S_3 \over S_2^2} \rangle  - 4 \; \gamma -1\over \alpha + 2\gamma+ \phi(1)}.
\label{r2infa}
\end{eqnarray}
 With $\langle S_3 \rangle \approx {1\over N^2}$, we have $\langle R_2 (\infty)\rangle= { 1 \over  2 \; |g'(1)| \; N_{\nu} } \; \langle {1\over S_2} \rangle $. 
 Our numerical analysis based on three different ensembles  indicates $\langle {1\over S_2} \rangle \sim N \log N$ for large $Y$ and large $N$ ((displayed in figure 5(d), details discussed in section IV). This implies $\langle R_2 (\infty)\rangle \sim \log N$, the expected limit for a maximally entangled state (as $\lambda_n \approx {1\over N}$ in large $Y$ limit, the logarithmic $R_2$ behavior is consistent with its expectation).

\section{Calculation of $\langle R_0 \rangle $: (stationary limit)}

Using the definition $R_0 \equiv - \sum_n \log \lambda_n$, its ensemble average can be given as 
\begin{eqnarray}
\langle R_0(Y) \rangle = -\int \left(\sum_n \log \lambda_n\right) \, P^{\beta}(\{\lambda\}; Y) \, D_{\lambda},
\label{r01}
\end{eqnarray}
Based on our numerical analysis of three different ensembles (details discussed in section IV), we find that, for $Y > o(1/N)$ range,  $\langle R_0 \rangle$ rapidly becomes almost constant with respect to $Y$ (numerical results displayed in fig. \ref{avgLoglmda}), approaching its stationary limit i.e $Y \to \infty$. As the ensemble in this limit corresponds to Stationary Wishart ensemble with unit trace with its eigenvalue {\it jpdf} given as 
$$P^{\beta}(\{\lambda\}) = C_{mn}^{\beta} \, \delta(\sum_n \lambda_n - 1) \, | \Delta _N (\{\lambda\}) |^{\beta} \, \Pi_j \lambda_j^{\omega}, $$
with $\omega \equiv \beta \nu - 1,$ and $ \Delta _N $ as the Vondermonde determinant, a substitution of the above in eq.(\ref{r01}) leads to $\langle R_0 \rangle $.
The integral in eq.(\ref{r01}) can be calculated by the \textit{auxilary Gamma function method}  discussed  \cite{apsk} for large $N$. We have
$$\langle R_0 \rangle = N \, \Psi\left(\frac{\beta N M}{2}\right) - \langle \log \lambda \rangle _{MP},$$ where, $\Psi(x) $ is the \textit{digamma function}, $M \geq N$, and $\langle \cdot \rangle _{MP} = \int _{\lambda_-} ^ {\lambda_+} \, \cdot \, \rho _{\lambda} \, d\lambda$, where $\rho _{\lambda}$ is the \textit{Marchenko-Pastur distribution},
\begin{eqnarray*}
 \rho _{\lambda} = \frac{\sqrt{(\lambda_+ - \lambda)(\lambda - \lambda_-)}}{\pi \beta \lambda},
\end{eqnarray*}
with $\lambda_{\pm} = \frac{\beta M}{2}\left(1 \pm \sqrt{\frac{N}{M}}\right)^2.$
Solving, we get,
\begin{equation}
  \langle R_0 \rangle = N \log N + N + (M- N) \, log \left(1-\frac{N}{M}\right) - \frac{1}{\beta M},
\end{equation}
Neglecting  the last two terms in large $N$ limit, we have,
\begin{equation}
\langle R_0 \rangle \approx N \log N + N .
\label{avgR0}
\end{equation}

\section{Derivation of relation(\ref{llr})}

Consider the function 
  \begin{equation}
    L = \sum _{n,m \atop n \neq m} \frac{\lambda_n \log \lambda_n}{\lambda_n - \lambda_m}.
\label{ll1}
  \end{equation}
with $\lambda_n$ as the eigenvalues of the ensemble under consideration.  
By rearrnaging the terms, $L$ can be rewritten as a series,  $L=\sum_k L_k$ with $L_k$ corresponding to $k^{th}$ order nearest neighbor contribution to 
$L$,
\begin{eqnarray}
    L_k &=& \sum_n \frac{\lambda_n \log \lambda_n - \lambda_{n+k} \log \lambda_{n+k}}{\lambda_n - \lambda_{n+k}}
\end{eqnarray}
 
We begin by approximating $L_1$ by writing $\lambda_{n+1} \approx \lambda_n + \Delta_n$ with  $\Delta_n$ as the local mean level spacing 
at $\lambda_n$. Using $\Delta_n  << \lambda_n$, one can write
$\log \lambda_{n+1} =  \log \lambda_n + \log(1 + \frac{\Delta_n}{\lambda_n})\approx \log \lambda_n + \frac{\Delta_n}{\lambda_n} - \frac{\Delta_n^2}{2 \lambda_n^2}$.
This in turn gives 
$$  \lambda_n \log \lambda_n - \lambda_{n+1} \log \lambda_{n+1}  \approx \Delta_n - \frac{\Delta_n^2}{2\lambda_n} + \Delta_n \log \lambda_n + \frac{\Delta_n^2}{\lambda_n} + \mathcal{O} (\Delta ^3)$$

Substituting the above in $L_1$ gives 
\begin{eqnarray}
L_1 &=& \sum_{n=1}^{N-1} (1 + \log \lambda_n) + \sum_{n=1}^{N-1} \frac{\Delta_n}{2 \lambda_n}\\
    &\approx& \sum_{n=1}^{N-1} (1 + \log \lambda_n).
  \end{eqnarray}
 
Similarly, considering next nearest neighbour term $L_2$, approximating $\lambda_{n+2} = \lambda_n + 2 \Delta_n$, and,  proceeding as in $L_1$ case, we have 
  \begin{eqnarray}
    L_2 &=& \sum_{n=1}^{N-2} (1 + \log \lambda_n) + \sum_{n=1}^{N-2} \frac{\Delta_n}{\lambda_n}\\
    &\approx& \sum_{n=1}^{N-2} (1 + \log \lambda_n).
  \end{eqnarray}
  Proceeding along the same route for $k >2$ i.e approximating 
  $\lambda_{n+k} \approx \lambda_n + k \; \Delta_n$, $L_k$ can be approximated as  $L_k \approx \sum_{n=1}^{N-k} (1 + \log \lambda_n)$.
Substitution of $L_k$ for all $k$ in eq.(\ref{ll1}), with $R_0 =- \sum_n \log \lambda_n$, then leads to 

  \begin{eqnarray}
    L &=& \sum_{i=1}^{N} \sum_{n=1}^{N-i} (1 + \log \lambda_n) \\
    &=& \frac{N (N-1)}{2} - \frac{(N-1)}{2} \langle R_0 \rangle.
  \end{eqnarray}

\begin{figure}[ht!]
\centering

\vspace{-1in}

\includegraphics[width=22cm,height=30 cm]{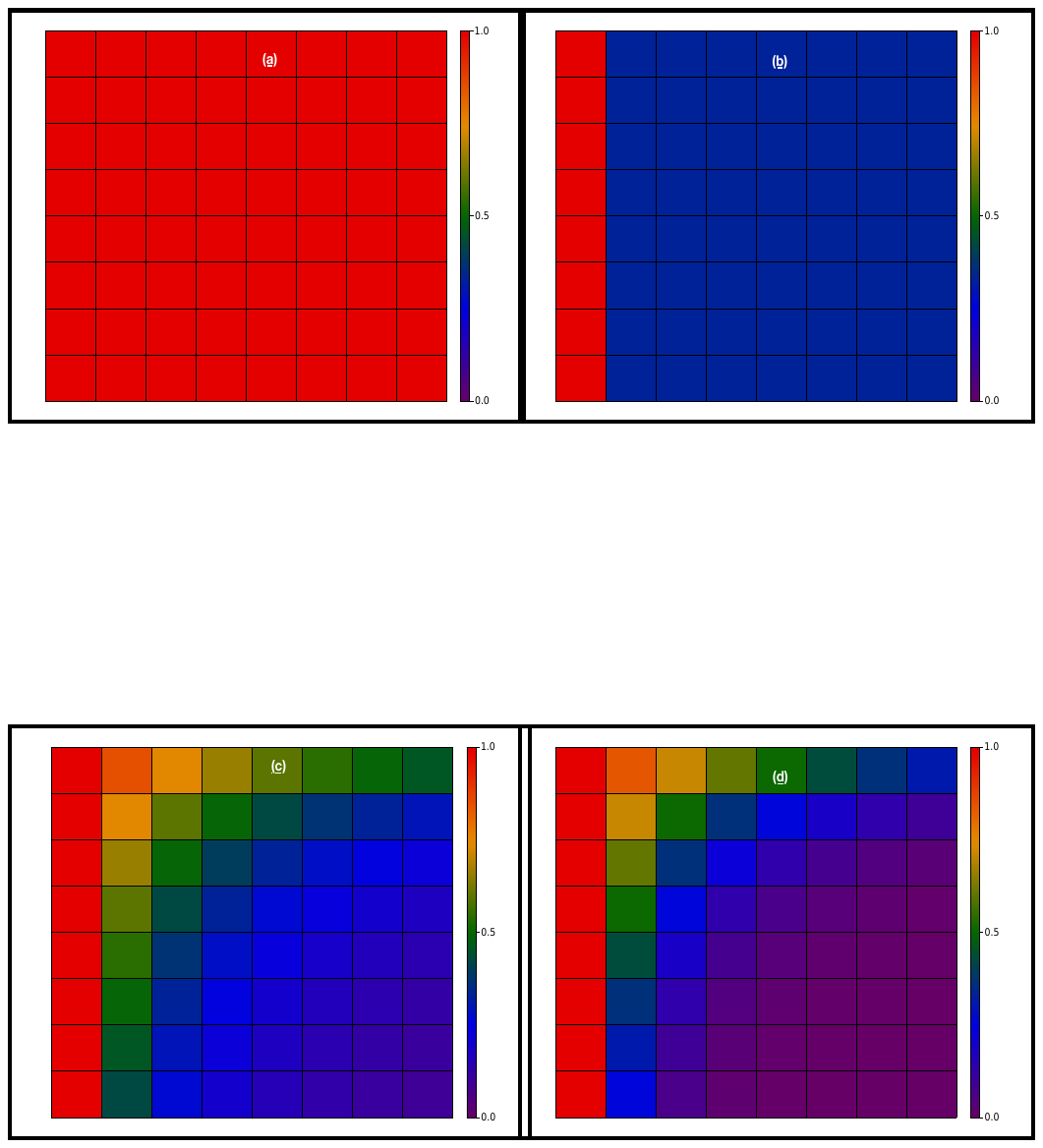}

\vspace{-3in}

\caption{{\bf  Schematic Representation of $C$-matrix:}  For visual guidance, the different protocols for variance distribution in the $C$-matrix used in our numerics are illustrated here: (a) Identical distribution (matrix taken from stationary Wishart ensemble) , (b) Constant variance except the first column (a matrix taken from ensemble eq.(\ref{vrp})) , (c) power-law-decay ((a matrix taken from \eqref{vpe})) and (d) exponential-decay along rows and columns (a matrix taken from (\ref{vee})) . The details for each case are given in section IV}
\end{figure}

\begin{figure}[ht!]
\centering

\includegraphics[width=22cm,height=40cm]{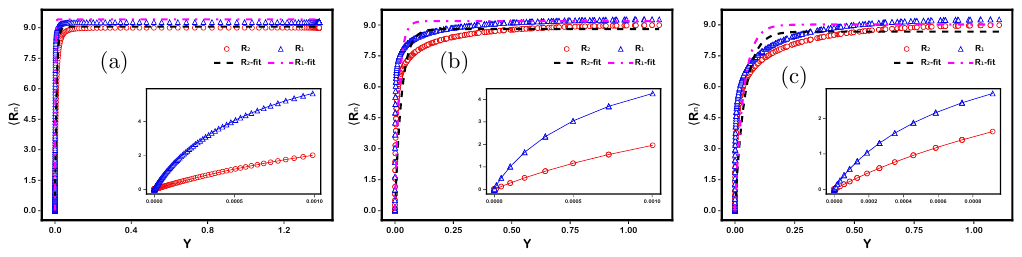}

\vspace{-11.9in}

\caption{{\bf  Growth of Average entanglement entropy for a large subsystem size $N = 1024$: } The figure displays a comparison of growth of average von Neumann entropy ($R_1$) and the second order R\'enyi entropy ($R_2$)   for ensembles with (a) constant variance, (b) power-law-decay, and (c) exponential-decay. We also fit the function $\langle R_{\alpha}(Y)\rangle = p_0 \, [1 - \exp(-p_1 \, Y)]$ for all the three cases, with the fit parameters respectively for $R_1$ and $R_2$ as (a) \{($p_0 = 9.41, \, p_1 = 509.06$), $p_0 = 9.05, \, p_1 = 123.04$\}, (b) \{($p_0 = 9.19, \, p_1 = 56.1$), $p_0 = 8.80, \, p_1 = 35.22$\}, (c) \{($p_0 =9.02, \, p_1 = 27.72$), $p_0 = 8.68, \, p_1 = 24.3$\}. The insets in each part highlight the different growth rate for $R_1$ and $R_2$. We note that $\beta=1$ and $\beta=2$ cases for each ensemble overlap with each other; this is consistent with our theoretical prediction.}
 \label{avgEntDynVN}
\end{figure}

\begin{figure}[ht!]
\centering

\includegraphics[width=20cm,height=30cm]{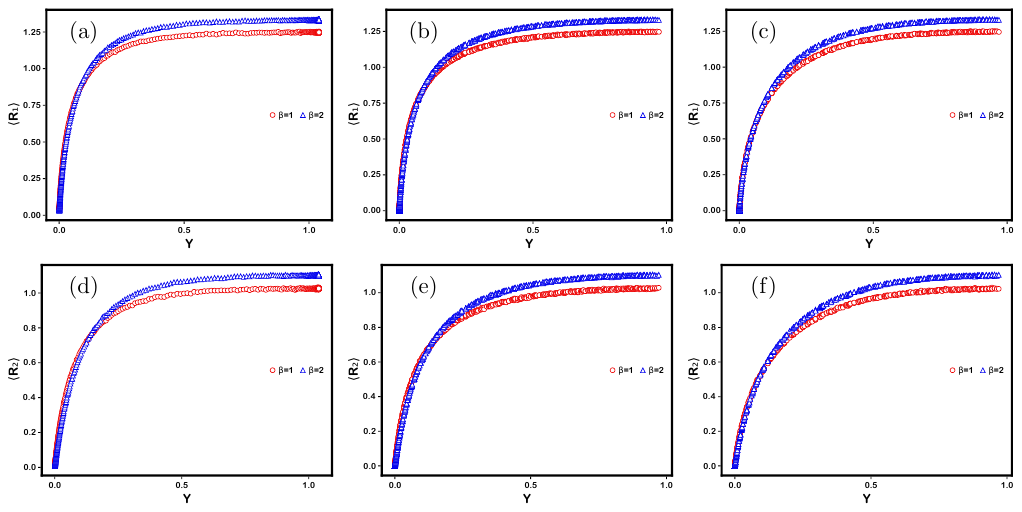}

\vspace{-6.8in}

  \caption{{\bf  Symmetry dependence of average entanglement entropy growth for a small subsystem size $N=4$: } A comparison of growth of average von Neumann entropy ($R_1$) for a typical state with and without time-reversal symmetry  is shown for three  Gaussian ensembles with (a) constant variance, (b) power-law-decay, and (c) exponential-decay, which is consistent with our analytical results. The same is shown for the second order R\`enyi entropy ($R_2$) in (d), (e), and (f).}  \label{compBetasR2}
\end{figure}

\begin{figure}[ht!]
\centering

\includegraphics[width=20cm,height=30cm]{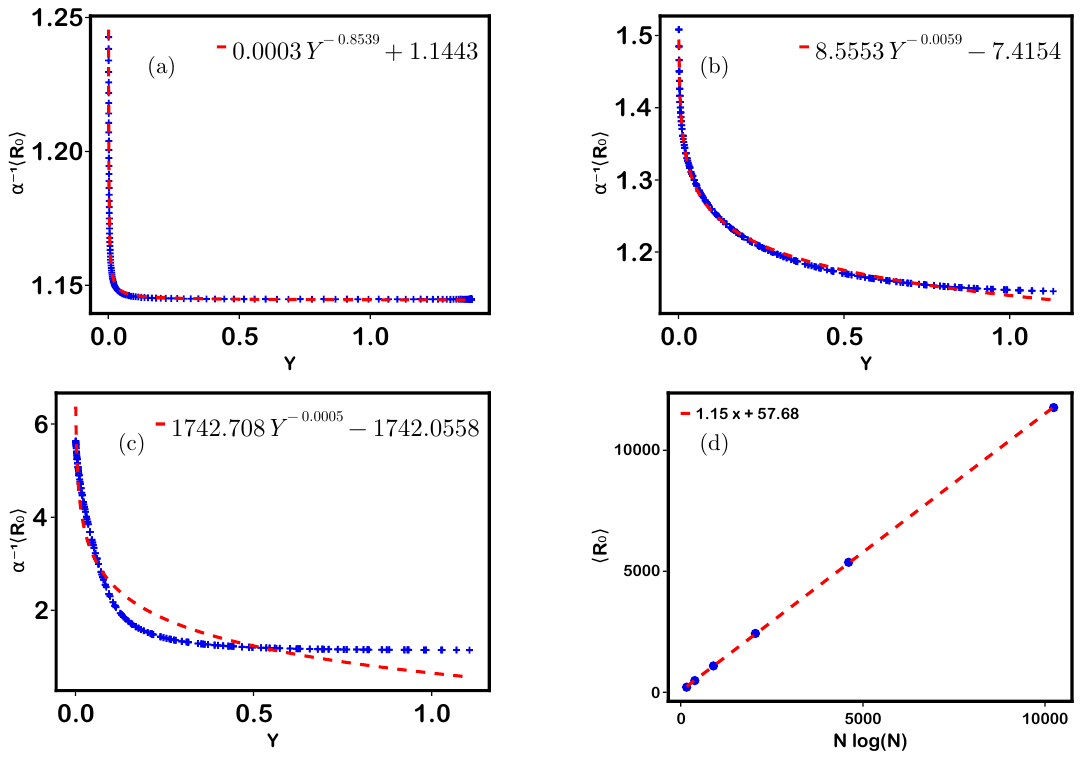}

\vspace{-5.1in}

  \caption{{\bf  $N$ and $Y$-dependence of $\langle R_0 \rangle$ :} Average behavior of  $R_0=-\sum_n \log \lambda_n$ (rescaled by $\alpha=N \log N$ for subsystem size $N=1024$, averaged over an ensemble of $1000$ matrices ) with respect to $Y$ is displayed for  ensemble with (a) constant variance, (b)  power-law decay, (c) exponential-decay (details in section IV). The part (d) of the figure confirms the expected behavior in large $N$ as well as large $Y$ limit: $\langle R_0 \rangle \propto N \log_2 N$; (here the result is displayed only for one of the ensembles, stationary limit of all ensembles being same). The dashed line correpond to  fits given in each figure}
\label{avgLoglmda}
\end{figure}

\begin{figure}[ht!]

\includegraphics[width=20cm,height=30cm]{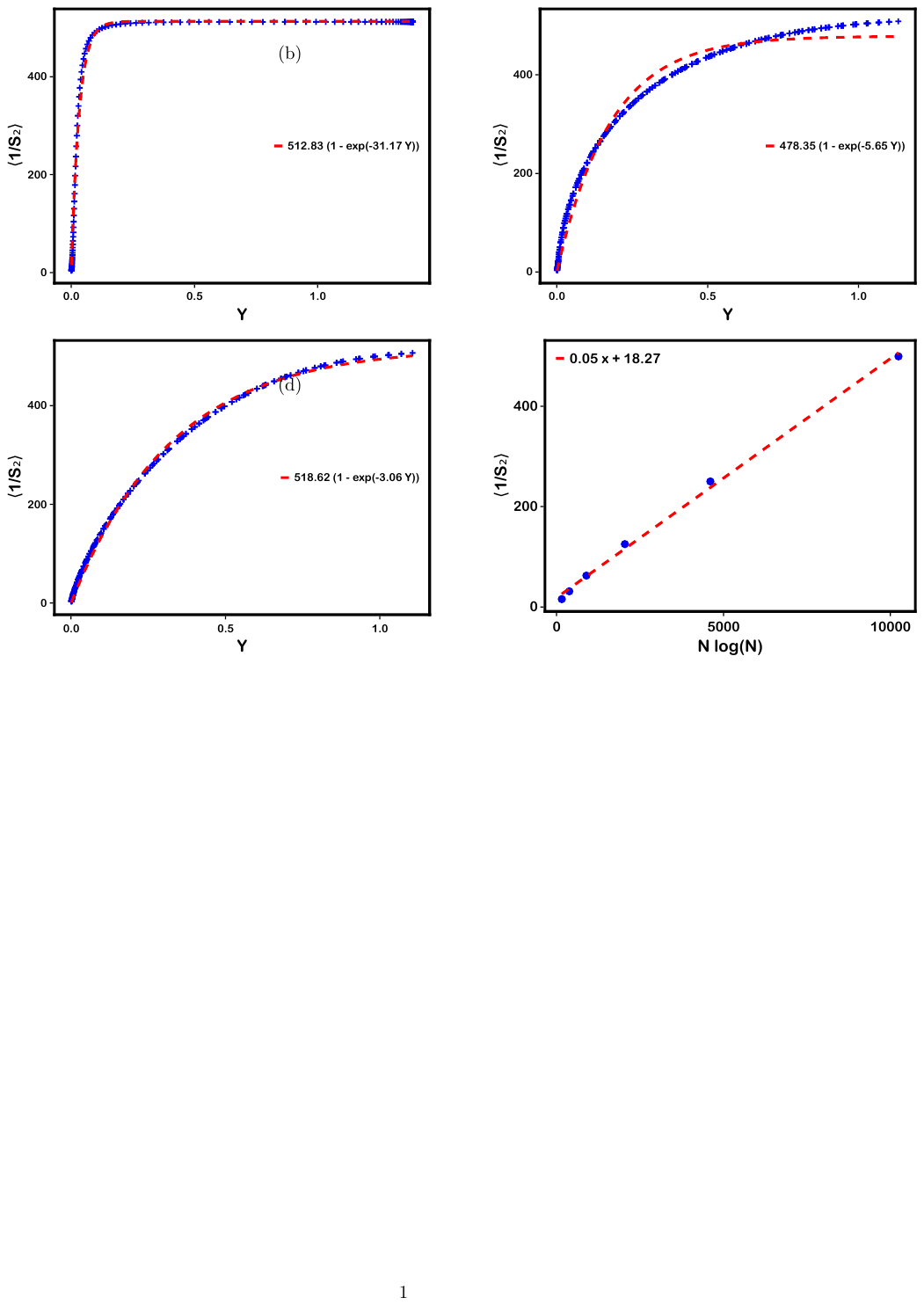}

\vspace{-5.1in}

  \caption{{\bf  $N$ and $Y$-dependence of $\langle {1\over S_2} \rangle$ :} 
  Average behavior of  ${1\over S_2}$ with $S_2=\sum_n \lambda_n^2$  (for subsystem size $N=1024$, averaged over an ensemble of $1000$ matrices ) with respect to $Y$ is displayed for ensemble with (a) constant variance, (b)  power-law decay, (c) exponential-decay (details in section IV).  The part (d) of the figure confirms the expected behavior in large $N$ as well as large $Y$ limit: $\langle {1\over S_2} \rangle \propto N \log_2 N$; (here again the display is only for one of the ensembles). The dashed line correpond to  fits given in each figure.}
  \label{avgInvS2}
\end{figure}

\begin{figure}[ht!]
\centering

\includegraphics[width=17cm,height=14cm]{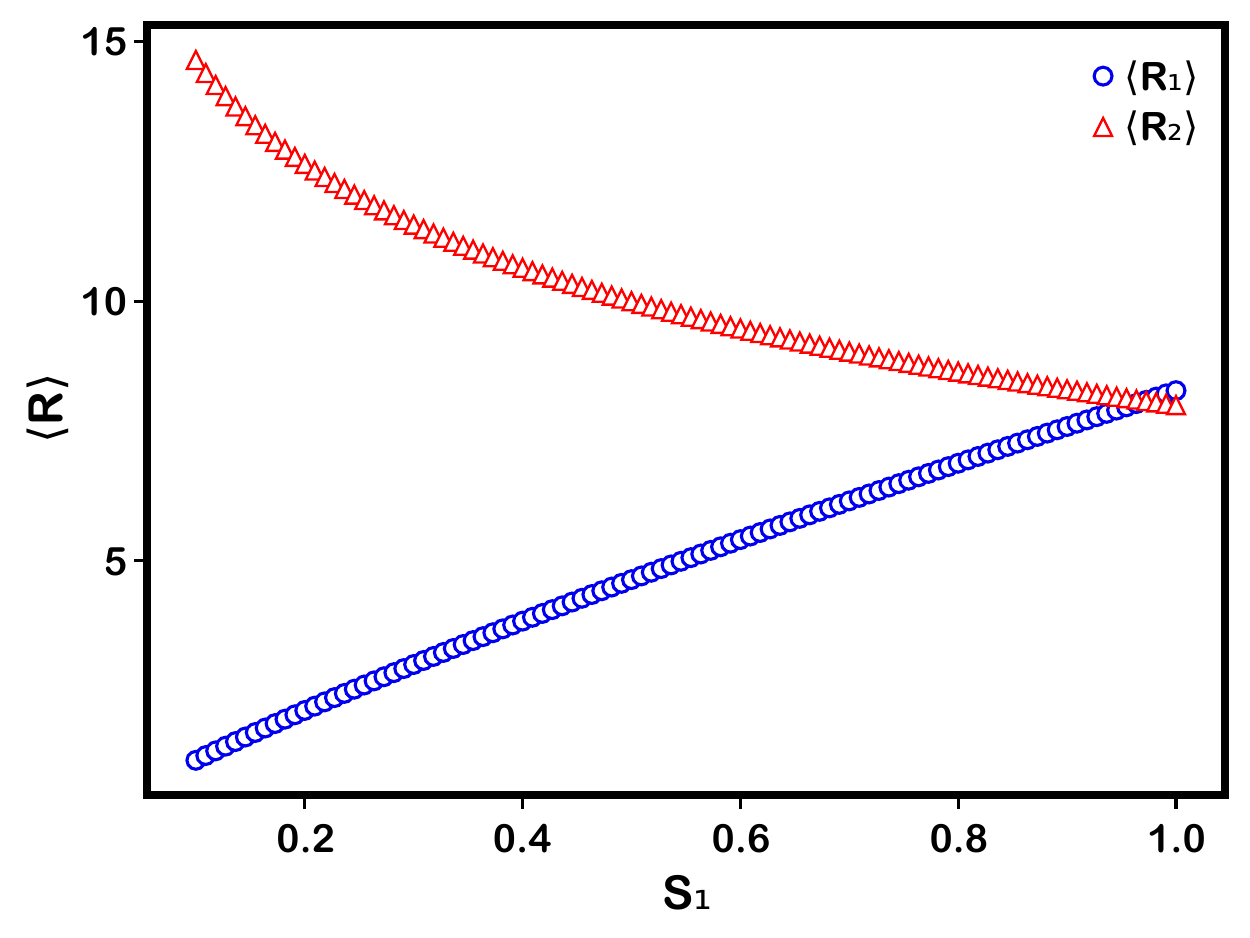}


\caption{{\bf  Trace constraint dependence of entanglement entropies:} Here the variation of $\langle R_1 \rangle$ and $\langle R_2 \rangle$ with $S_1=\sum_n \lambda_n $ is shown for the standard Wishart ensemble, for $N = 512$. As clear from the display, ${\partial \langle R_1 \rangle \over \partial S_1} >0)$ and   ${\partial \langle R_2 \rangle \over \partial S_1} < 0)$.  This infomation is used in section III. A and section III.B.}
\label{r2r1vsS1}
\end{figure}


\newpage

\begin{thebibliography}{10}

\bibitem{zhh}
  K. {\.Z}yczkowski, P. Horodecki, M. Horodecki, and R. Horodecki,
  Phys. Rev. A 65, 012101 (2002).


\bibitem{nahum1} A. Nahum, J. Ruhman, S. Vijay, J. Haah, Physical Review X, 7, 031016 (2017)


 \bibitem{vhbr}
 L. Vidmar, L. Hackl, E. Bianchi, and M. Rigol, Phys. Rev. Lett. 121, 220602, (2018).
 
  \bibitem{nh}
R. Nandkishore and D. A. Huse,  Annu. Rev. Condens. Matter Phys. 6, 15, (2015).

\bibitem{bgl}
 G. Bentsen, Y. Gu, and A. Lucas, PNAS, 116, 6689, (2019).

\bibitem{ha}
 W. W. Ho and D. A. Abanin,  Phys. Rev. B, 95,  094302, (2017).

  \bibitem{zl}
  T. Zhou and A.W.W.Ludwig, Phys. Rev. Research, 2, 033020, (2020).


  \bibitem{rpv}
  T. Rakovszky, F. Pollmann and C.W.von Keyserlingk, Phys. Rev. Lett. 122, 250602,(2019).

  \bibitem{rpk27}
 T. Zhou and A. Nahum, Phys. Rev. B 99, 174205 (2019). 
 
\bibitem{rpk28}
B. Bertini, P. Kos, and T. Prosen, Phys. Rev. X 9, 021033 (2019).


  \bibitem{rpk15}
 V. Alba and P. Calabrese, SciPost Phys. 4, 17 (2018).

 \bibitem{rpk19}
 M. Mezei, Phys. Rev. D 98, 106025 (2018).

 \bibitem{rpk24}
 M. Mestyán, V. Alba, and P. Calabrese, J Stat. Mech. (2018), 083104.
 
 \bibitem{rpk25}
C. W. von Keyserlingk, T. Rakovszky, F. Pollmann, and S. L. Sondhi, Phys. Rev. X 8, 021013 (2018).

\bibitem{rpk26}
A. Nahum, S. Vijay, and J. Haah, Phys. Rev. X 8, 021014 (2018).

  \bibitem{nhkw}
  Y. Nakata, C. Hirche, M. Koashi and A. Winter, Phys. Rev. X7, 021006, (2017).

  \bibitem{rpk14}
  V. Alba and P. Calabrese, Proc. Natl. Acad. Sci. U.S.A. 114,
  7947 (2017).

  \bibitem{rpk17}
  A. Nahum, J. Ruhman, S. Vijay, and J. Haah, Phys. Rev. X7, 031016 (2017).

  \bibitem{rpk20} 
 W. W. Ho and D. A. Abanin, Phys. Rev. B 95, 094302, (2017).

\bibitem{rpk21}
 M. Mezei and D. Stanford, J. High Energy Phys. 17(2017) 65.

 \bibitem{rpk23}
V. Alba and P. Calabrese, J. Stat. Mech. (2017) 113105. 

  \bibitem{puz}  
 Z. Pucha{\l}a, {\L}. Pawela, K. {\.Z}yczkowski, Phys. Rev. A 93, 061221 (2016).

  \bibitem{efg}
  J. Eisert, M. Friesdorf and C. Gogolin, Nature Physics, 11, 124, (2015).

  \bibitem{imptlrg}
  R. Islam, R. Ma, P.M. Press, M. Eric Tai, A. Lukin, M. Rispoli and M. Greiner, Nature, 528, 77, (2015).

  \bibitem{hp}
  P. Hayden and J. Preskill, Journal of High Energy Physics, 09, 120, (2007).

  \bibitem{ss}
  Y. Sekino and L. Susskind, Journal of High Energy Physics, 10, 065, (2008).

 
  \bibitem{arul}
  J. N. Bandyopadhyay and A. Lakshminarayan, Phys. Rev. Lett., 89, 060402, (2002).



\bibitem{rpk12}
P. Calabrese and J. Cardy, J. Stat. Mech. (2005) P04010. 

\bibitem{rpk13}
 P. Calabrese and J. Cardy, J. Stat. Mech. (2007) P10004. 
 

\bibitem{rpk16} 
 H. Kim and D.A. Huse, Phys. Rev. Lett. 111, 127205
(2013).

\bibitem{jhn}
C. Jonay, D. A. Huse, and A. Nahum, arXiv:1803.00089. 
 
\bibitem{rpk22}
M. Fagotti and P. Calabrese, Phys. Rev. A 78, 010306(R)
(2008).

\bibitem{apsk} S. Kumar, S. \& A. Pandey,  JPhys. A: Math. \& Theo. 44, 445301 (2011).

 \bibitem{ptche}
  T. Mondal and P.Shukla, Phys Rev. E 102, 032131 (2020)
  
  
 \bibitem{pche}
P.Shukla, J. Phys. A: Math. Theor. 54 275001, (2021).

\bibitem{sp}
S. Kumar and A. Pandey, Ann. Phys. 326, 1877, (2011).

\bibitem{pslg}
P. Shukla, J. Phys. A: Math. Theor 50, 435003 (2017).


\bibitem{berry}
M.V. Berry, J. Phys. A: Math. Gen. 10, 2083 (1977).

\bibitem{page}
D. Page, Phys.Rev.Lett.71:1291-1294,1993.
r
\bibitem{virmani2000ordering}Virmani, S. \& Plenio,  Physics Letters A, 268, 31-34 (2000).


\end{thebibliography}
\end{document}